\documentclass{article}
\providecommand{\undertitle}{}

\usepackage{arxiv}
\usepackage{amsmath} 
\usepackage[utf8]{inputenc} 
\usepackage[T1]{fontenc}    
\usepackage{hyperref}       
\usepackage{url}            
\usepackage{booktabs}       
\usepackage{amsfonts}       
\usepackage{nicefrac}       
\usepackage{microtype}      
\usepackage{lipsum}		
\usepackage{graphicx}
\usepackage{natbib}
\usepackage{doi}
\usepackage{fancyhdr} 
\newcommand{\headeright}{}

\title{Modeling Time-Dependent Responses of Optical Compressors with Selective State Space Models}

\date{} 					

\author{ {\hspace{1mm}Riccardo Simionato} \\
	Department of Musicology\\
	University of Oslo\\
	Oslo, Norway \\
	\texttt{riccardo.simionato@imv.uio.no} \\
	\And
	{\hspace{1mm}Stefano Fasciani} \\
	Department of Musicology\\
	University of Oslo\\
	Oslo, Norway \\
	\texttt{stefano.fasciani@imv.uio.no} \\
}

\hypersetup{
pdftitle={A template for the arxiv style},
pdfsubject={q-bio.NC, q-bio.QM},
pdfauthor={David S.~Hippocampus, Elias D.~Striatum},
pdfkeywords={First keyword, Second keyword, More},
}

\begin{document}
\maketitle
\begin{abstract}
This paper presents a method for modeling optical dynamic range compressors using deep neural networks with Selective State Space models. The proposed approach surpasses previous methods based on recurrent layers by employing a Selective State Space block to encode the input audio. It features a refined technique integrating Feature-wise Linear Modulation and Gated Linear Units to adjust the network dynamically, conditioning the compression's attack and release phases according to external parameters. The proposed architecture is well-suited for low-latency and real-time applications, crucial in live audio processing. The method has been validated on the analog optical compressors TubeTech CL 1B and Teletronix LA-2A, which possess distinct characteristics. Evaluation is performed using quantitative metrics and subjective listening tests, comparing the proposed method with other state-of-the-art models. Results show that our black-box modeling methods outperform all others, achieving accurate emulation of the compression process for both seen and unseen settings during training. We further show a correlation between this accuracy and the sampling density of the control parameters in the dataset and identify settings with fast attack and slow release as the most challenging to emulate.
\end{abstract}

\section{INTRODUCTION}
The emulation of analog musical devices constitutes a significant portion of consumer-grade music production and performance software. The nonlinear characteristics of their electrical and electronic components give analog audio effects a distinctive timbre and sound quality that is highly sought after by both music professionals and enthusiasts. Accurate modeling of these components has become an active area of research and development, commonly known as Virtual Analog (VA) modeling.

Modeling of analog effects typically employs one of two main approaches: physics-based or data-driven, each with distinct characteristics. The physics-based method depends on mathematical descriptions of physical processes, whereas the data-driven strategy leverages data collected from the devices themselves. Physics-based modeling involves deriving and solving parametric mathematical equations that accurately describe an effect's behavior and allow for manipulating the effect's control parameters. This task is nontrivial; moreover, without simplifying assumptions, physics-based approaches can be computationally complex and may not be suitable for consumer-grade products. Among physics-based approaches, wave digital filter (WDF) techniques have gained particular popularity; these have been utilized for modeling vacuum tube-based circuits ~\citep{dunkel2016fender}, envelope filter guitar effect circuit consisting of operational transconductance amplifiers ~\citep{bogason2017modeling}, cascaded amplifiers ~\citep{zhang2018real}, a booster guitar pedal based on field-effect transistors ~\citep{bogason2018modeling}, ring modulators ~\citep{bernardini2018wave}, and the phaser MXR 90 ~\citep{giampiccolo2024wave}. 

Conversely, data-driven approaches derive digital signal processing algorithms by analyzing device measurements, aiming to replicate the input-output relationship rather than the underlying physical phenomena. These approaches often encounter limitations in terms of flexibility. Since models are obtained from a limited set of measurements, and there is a theoretically infinite combination of possible inputs, they may either be limited to operating exclusively within scenarios represented in the measurements, where sufficient accuracy can be verified, or they can be used to predict outputs for input combinations unseen in the measurement data, where accuracy may not be absolutely guaranteed, especially when inferring values outside the range of the measurement data. Common examples of data-driven approaches include block-oriented models such as the Hammerstein and Wiener models, as well as their various combinations ~\citep{schoukens2017identification}. These models have been utilized for guitar distortion pedals ~\citep{eichas2015block}, guitar amplifiers ~\citep{schuck2016audio, eichas2017block}, distortion circuits ~\citep{eichas2016black}, and guitar overdrive effect pedals ~\citep{koper2020taming}.

Recent advancements in neural networks have provided novel tools to mitigate the drawbacks of existing data-driven methodology. In particular, neural networks demonstrate generalization capabilities that can surmount classical data-driven algorithms. The first successful uses of artificial neural networks (ANNs) to emulate analog effects were in modeling amplifiers and distortion effects ~\citep{mendoza2005emulating, covert2013vacuum, damskagg2019real, wright2019real}. These works have highlighted key findings regarding the viability of artificial neural networks for Virtual Analog (VA) modeling: the models are perceptually accurate; the networks can be conditioned with the effect's parameters, allowing for dynamic changes in the sound alteration process; accurate models can be achieved when processing raw audio samples, with the networks satisfying real-time computational constraints on consumer-grade signal processing hardware; and the networks can process single or small blocks of audio samples, thus ensuring minimal or negligible latency suitable for live audio applications.

More complex types of audio effects have also been successfully modeled using ANN-based modeling techniques. This includes time-based effects such as delay ~\citep{mikkonen2023neural} and reverb ~\citep{ramirez2020modeling}, modulation effects like chorus ~\citep{wright2021neural}, and dynamic effects like compressors ~\citep{steinmetz2021efficient, simionato2022deep, simionato2023fully}. These effects typically involve a sound alteration process with complex and often lengthy temporal dependencies between the input and output signals. However, at least in their basic form, artificial neural networks are memoryless computational structures. Common variations incorporating memory, such as recurrent networks, have shown significant limitations in modeling long temporal dependencies ~\citep{bai2018empirical, simionato2022deep}. As a result, successful attempts to model these complex audio effects with ANNs have utilized large-scale networks that process audio signals in large blocks of samples. Consequently, these models generally have poor suitability for actual live audio applications due to their excessive latency or high computational complexity.

This article introduces a method for the neural modeling of dynamic range compression (DRC), focusing on analog optical compressors. DRC is a nonlinear effect that reduces the dynamic range of the input signal ~\citep{giannoulis2012digital}. DRC is widespread in music production and live audio, typically employed to decrease the volume of louder segments, enhance quieter parts, or control transient peaks. Another common application is the `sidechain' mode, where an external signal triggers the compression of the input signal based on its amplitude, often used to prevent masking in concurrent sounds, such as a kick drum and bass or a lead vocal and background elements. In radio broadcasting, a compressor can automatically reduce the music volume when someone speaks.

Although not all compressors feature the same type and number of parameters, the function of DRC is characterized by four main controllable quantities: threshold, compression ratio, attack time, and release time. The threshold sets the amplitude level above which the compressor's gain reduction mechanism is activated, and the ratio determines the degree of compression applied. Attack and release times dictate how quickly the gain reduction begins and ends after exceeding the threshold. DRC applies a time-dependent and level-dependent gain to the input signal.

In this work, we consider optical compressors. When the input signal exceeds the threshold, the compressor diverts some of the voltage to the gain-reducing optical circuit. The signal voltage drives a light filament, which heats up and begins emitting light that progressively increases. The emitted light causes a light-sensitive variable resistor to increase its resistance, thereby reducing the output gain. This process starts immediately after the threshold is crossed, but there can be a lag in gain reduction due to the time it takes for the filament to heat up and emit sufficient light. This leads to a characteristically smooth response, often seen in optical analog compressors. Furthermore, this response time depends on the frequency and is also affected by previous heating cycles; hence, the input-output relationship can be quite complex, with significant dependencies on earlier parts of the input signal. These characteristics are challenging to capture with physical modeling techniques.

Building upon our previous contributions to analog DRC modeling, we have enhanced the neural architecture incorporating the Selective State Space (S6) model~\citep{gu2023mamba} while preserving the original design's low latency and small computational costs. State space models have demonstrated their potential in sequence-to-sequence modeling tasks and, more recently, in capturing the significant temporal characteristics of analog effects ~\citep{simionato2024comparative}. 

We further enhanced the black-box modeling approach by integrating all the user-controllable parameters using Feature-wise Linear Modulation (FiLM) ~\citep{perez2018film} and Gated Linear Units (GLU) ~\citep{dauphin2017language} methods to condition the networks. Following the proposal in ~\citep{simionato2024conditioning}, we placed the conditioning layer after the State Space layer. We also introduced a second layer after the conditioning block to provide additional handling of nonlinearities and time dependencies. Given that the DRC's control parameters govern various aspects of its operation, we divided the conditioning vector into two parts: one addressing dynamics and another focusing on timing characteristics, for which we employed the Temporal FiLM method ~\citep{birnbaum2019temporal}. Furthermore, we confirmed the efficacy of using the window technique, which aids the network in inferring the correct output sample by providing auxiliary past input information. In this context, we leveraged the spectral magnitude information from the input signal to facilitate the conditioning process.

We compare the proposed modeling method with other approaches based on the same architecture but featuring different types of layers: Long Short-Term Memory (LSTM) ~\citep{hochreiter1997long} and Structured State Space (S4D) model ~\citep{gu2022parameterization}. We evaluate the models' overall learning abilities and the ability of the models to predict compression for unseen combinations of conditioning parameters. We investigate the impact of the control parameters' dataset sampling resolution on the network's ability to interpolate between values seen during training. Finally, listening tests are conducted to evaluate the perceptual quality of the models subjectively.

\section{BACKGROUND}\label{sec:VA}

In the last two decades, machine learning techniques, particularly artificial neural networks (ANNs), have been extensively employed for modeling analog audio effects. These techniques have been adopted as a black-box approach to emulate audio effects processing, manipulating raw audio samples in a manner similar to digital signal processing (DSP) methods.

In one of the early attempts, Mendoza ~\citep{mendoza2005emulating} utilized a multilayer feedforward network to emulate both a high-pass filter and the Ibanez Tube-Screamer distortion effect. Although the network did not achieve satisfactory results when trained on the high-pass filter in the frequency domain, training in the time domain for the distortion effect showcased the potential of ANNs to learn and accurately replicate audio distortion effects. Feedforward networks are the simplest type of ANN. They do not have mechanisms to encode past information from the input data or to remember the system's state. To address this limitation, particularly when the modeling problem involves input-output temporal dependencies, the network's input is fed with a segment of the input signal. This segment includes already-seen past samples and is often much longer than the generated output, ensuring that each input-output pair used to train the network adequately represents such time dependencies. This approach is suitable for very short-time dependencies or may result in impractically high input-output latency for live audio applications.

Damskägg et al. proposed to utilize one-dimensional convolutional-based networks to model the preamplifier circuit of the Fender Bassman 56F-A vacuum-tube amplifier ~\citep{damskagg2019deep}. This research demonstrated that Convolutional Neural Networks (CNNs) can significantly improve the accuracy of audio effect emulation compared to fully connected (FC) feedforward networks when variable control parameters are incorporated into the model. Employing the same architecture, models of the Ibanez Tube Screamer, the Boss DS-1, and the Electro-Harmonix Big Muff Pi distortion pedals have also been developed ~\citep{damskagg2019real}.

Similarly to FC feedforward networks, CNN's ability to track the past information in the signal depends on the size of the receptive field; a larger receptive field necessitates a greater number of input past samples to predict a smaller set of more recent output samples. To overcome this limitation, Recurrent Neural Networks (RNN) have also been explored to model distortion effects. Specifically, RNN-based models of devices like the 4W Vox AC4TV vacuum-tube amplifier ~\citep{covert2013vacuum}, the Ibanez Tube Screamer, the Boss DS-1, and the Electro-Harmonix Big Muff Pi distortion pedals ~\citep{wright2019real} have been developed and compared to their respective CNN-based models. These studies employ Long Short-Term Memory (LSTM) and Gated Recurrent Units (GRU) networks as the recurrent layers. Results have shown that RNNs can achieve accuracy comparable to CNNs while offering a significantly lower computational cost for the inference and substantially reduced input-output latency, as they require only the current input sample to predict the current output sample.

Steinmetz et al., in what appears to be the earliest attempt to model a DRC ~\citep{steinmetz2021efficient}, proposed the use of Temporal Convolutional Networks (TCNs). In this work, the modeled device is the Teletronix LA-2A leveling amplifier, which features an optical gain reduction mechanism - challenges associated with its modeling were discussed in the introductory section. The model operates at a sampling rate of $44.1$~kHz. The TCN employs dilated convolutional layers with rapidly growing dilation factors ranging from $2^2$ to $2^{10}$. As the dilation factor increases, so does the receptive field, enabling the network to handle long temporal dependencies without adding more layers. This approach led to the development of a model with relatively low computational demands; however, it still required significantly long segments of the input signal, which is unfavorable for latency. The various models used in comparative experiments exhibit input-output latencies between $100$ and $1000$ ms, corresponding to their respective receptive field lengths. The model with a receptive field of $300$ ms demonstrated the best accuracy. 

To enhance past signal information encoding in networks without increasing layer size, we proposed an Encoder-Decoder (ED) architecture utilized to model the optical compressor TubeTech CL 1B ~\citep{simionato2022deep}. The network comprises two LSTM layers: an encoder working with a brief history of input samples and compressor settings, and a decoder for outputting predictions per input sample. The encoder continuously computes its internal states, which are then utilized by the decoder's LSTM layer as initial states. This transfer facilitates the decoder in generating accurate predictions by making use of a condensed history of past inputs. This ED approach notably outperformed traditional LSTMs and FC networks in accuracy, despite having a similar count of trainable parameters. Operating at a sampling rate of $48$ kHz, the model incorporates only two variable control parameters: threshold and ratio. This study examined networks with input segments ranging from $2$ to $16$ audio samples, resulting in a worst-case latency of $0.33$ ms. The results show a positive correlation between input segment size and accuracy.

Subsequently, we augmented the model by including attack and release times as variable parameters with an enhanced architecture, where a convolutional layer replaced the encoder's LSTM ~\citep{simionato2023fully}. This modification led to improved accuracy, albeit with a minor increase in the computational cost. However, this approach eliminated the continuous updating of internal states across sequences; states are recalculated at each iteration based on the number of the encoder's input samples. We experimented with up to $64$ samples, bounding the worst-case latency to $1.33$ ms. The network size was kept small to limit computational complexity. 

The best-performing model among those developed was compared against the top TCN model from ~\citep{steinmetz2021efficient} using our TubeTech CL 1B dataset and their Teletronix LA-2A dataset. Results showed that our model exhibited better accuracy in the CL 1B case. In comparison, the accuracy was slightly worse but comparable to that of the TCN in the LA-2A case despite the TCN model using significantly larger input segments. However, when the TCN was adapted to work with shorter input segments—thereby matching the input-output latency of our model—its accuracy was significantly inferior to that of our ED model. Moreover, the CL1B dataset included cases with significant compression at low thresholds (e.g., $-30$, $-40$ dBu), which results in audible artifacts due to minor signal amplitude mismatching at the edges of consecutive output audio segments. These artifacts were observed in both the ED and TCN models. However, the models exhibit limitations in scenarios of heavy compression, where they fail to replicate the abrupt and rapid gain reduction that typically occurs with onsets. Rapid attacks, when combined with heavy compression, are particularly challenging to model, as are extended-release times, which can span up to $10$ seconds in the case of the CL 1B.

A gray-box model, which combines information about the device's functioning with automatic parameter estimation, was proposed for the Teletronix LA-2A optical leveling amplifier ~\citep{graycomp}. The model incorporates the structure of a traditional digital compressor and predicts the parameters of the compression curve, including the threshold, ratio, and knee, using FC- and RNN-based networks. Similar techniques are employed to predict the timing parameters that define the attack and release envelopes of the equations designed and used to describe the compressor. This approach leverages domain knowledge to help generate audio with fewer artifacts and reduce the number of trainable parameters. 

The accuracy of this approach, however, is constrained by the inherent approximations present in the equations representing the system. In this study, for example, timing effects are modeled using three methods: the first assumes equal attack and release times, independent of the input signal; the second assumes varying attack and release times, but still not in response to the input signal; and the third assumes attack and release times are equal but do vary based on the input signal. The latter two methods demonstrated better performance. However, all these scenarios fail to accurately represent the actual timing behavior of the device, as, in reality, attack and release times may vary independently of each other and are influenced by the signal's energy. Consequently, due to these approximations, the model is inaccurately biased, which compromises its precision. On the other hand, the model brings the advantage of requiring $10\%$ of the number of operations needed by a black-box GRU model with $32$ units. In addition, the gray-box model does not require long input segments to achieve greater accuracy and, as a result, does not introduce added latency.

In most optical compressors, such as the LA-2A and CL1B, formalizing the variable attack and release behavior with mathematical expressions is not straightforward, which makes the use of black-box design potentially advantageous. The complex nonlinear and time-dependent behavior of compressors presents a significant challenge for traditional RNN and CNN architectures. Recent advances in sequence-to-sequence modeling have introduced State Space models (SSMs) that rely on a continuous-time representation of the state space formulation ~\citep{gu2022efficiently}. These models have demonstrated promising results, especially with long sequences. Similar to RNNs, they retain state information while addressing the vanishing memory problem by eliminating nonlinearity in the computation of states.

\section{PROPOSED MODEL}\label{sec:architecture}

This work proposes using Selective State Space (S6) networks to model optical dynamic range compressors. The architecture we have designed offers advantages due to its superior capability for capturing temporal dependencies in the input-output signals. As previously mentioned, in the S6 models, the matrix governing the computation of the internal states depends on the input signal. This represents an improvement over earlier efforts based on the S4D model in audio effect modeling, where the matrix does not depend on the input. This approach is particularly beneficial for capturing the complex interplay between the input signal and gain reduction—a relationship dominated by the coupling of light-emitting and light-sensing elements—and is especially useful for accurately modeling the characteristically slow attack and release times of optical compression. Our investigation is restricted to experiments with relatively small networks with limited computational complexity and latency, making them suitable for consumer-grade systems and live audio applications. We assess modeling accuracy by comparing the proposed Selective State Space-based architecture against the state-of-the-art models, including those based on LSTM, ED, S4D, and TCN.

\subsection{Network Architecture}\label{sec:nn}

\begin{figure}[h]%
\centering
\includegraphics[width=0.2\textwidth]{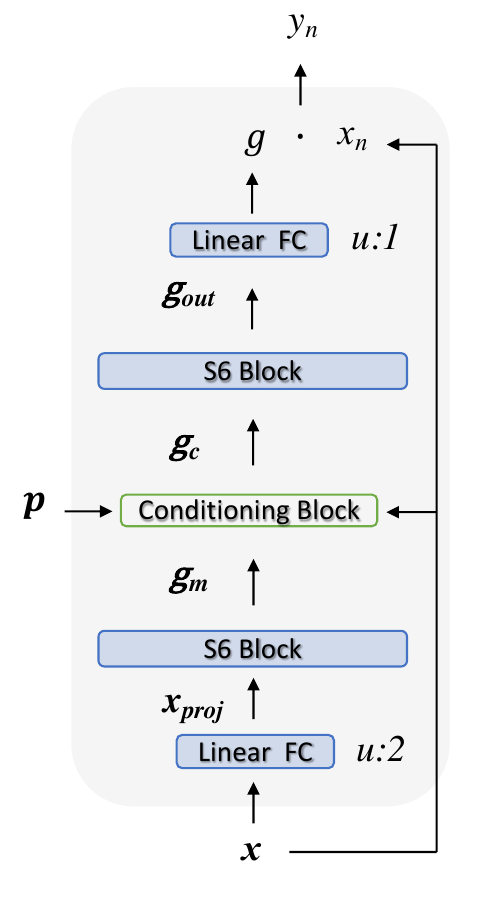}
\caption{Proposed architecture: the input $\boldsymbol{x}$, which contains the current and past samples, is fed to a linear FC layer and subsequently to a S6 block (detailed in Figure~\ref{fig:Ma}). The resulting vector passes through the conditioning layer and another identical S6 block before reaching the output layer, which is a linear FC layer with one unit. The number of units (u) is indicated next to each layer. The output is a coefficient $g$ which multiplied by the current input sample $x_n$ yields the current output sample $y_n$ }\label{fig:arch}
\end{figure}

The architecture we propose for modeling analog optical compression is centered on Selective State Space (S6) layers, as illustrated in Figure~\ref{fig:arch}. It includes a linear FC layer that processes and compresses the input, which then feeds into an S6 block. The conditioning layers process this block's output based on the values of the compression parameters $\boldsymbol{p}$, before passing its output through a second, identical S6 block. The output layer consists of a single unit that predicts a coefficient. When multiplied by the input sample, this coefficient produces the output sample. 

The architecture generates one output audio sample per inference cycle, taking as input a buffer of the $64$ most recent input samples. This implies that with each inference cycle, a new input sample enters the buffer, while the oldest is discarded. Therefore, the set of the $64$ most recent input samples constitutes the network input, which, after initial compression by the FC layer, is used to update the states in the S6 layers. This approach aids the network in making predictions while maintaining a stateful design using truncated backpropagation through time. Predicting one sample at a time helps minimize the audible artifacts commonly associated with machine-learning models of audio effects, which arise from slight amplitude mismatches at the boundaries of consecutive output segments. 

The mathematical description of the S6 layer is the following:
\begin{align}
     \boldsymbol{h}_n &= \boldsymbol{A} \boldsymbol{h}_{n-1} + \boldsymbol{B} \boldsymbol{j}_n \nonumber \\
     \boldsymbol{o}_n &= \boldsymbol{C} \boldsymbol{h}_n + \boldsymbol{D}\boldsymbol{j}_n \label{eq:ssm}
\end{align}
where $\boldsymbol{j}_n$ and $\boldsymbol{o}_n$ are the input and output vectors at time $n$. $\boldsymbol{A}_{N\times N}$, $\boldsymbol{B}_{N \times M}$, $\boldsymbol{C}_{M \times N}$, and $\boldsymbol{D}_{M \times M}$ are complex or real-valued matrices expressing linear mappings between $\boldsymbol{j}_n$ and $\boldsymbol{h}_n$ the states of the system. The layer captures the long-range dependencies using the state matrix $\boldsymbol{A}$. This matrix encodes all the past input history in $\boldsymbol{h}_n$, finding a map from an M-dimensional space $\boldsymbol{j}_n$ to a N-dimensional space $\boldsymbol{h}_n$ that represents the compression of the input's past. In the S6 layer, $\boldsymbol{B}$ and $\boldsymbol{C}$ are dependent on the input and computed using a linear FC layer, while $\boldsymbol{A}$ and $\boldsymbol{D}$'s parameters are independent of the input and learned during the training process. 
\begin{figure}[h]%
\centering
\includegraphics[width=0.3\textwidth]{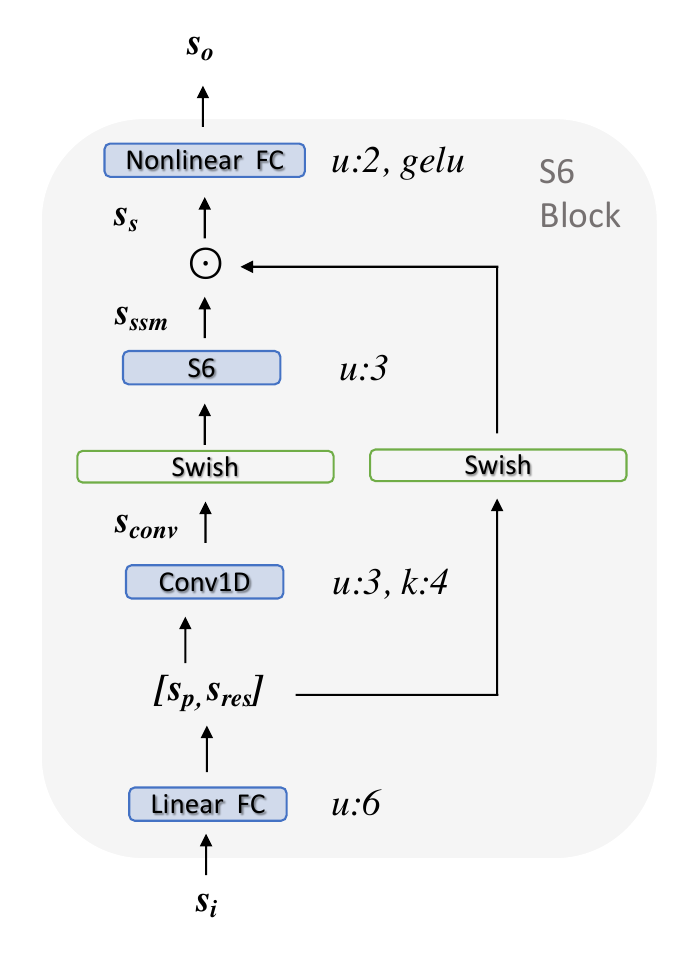}
\caption{Internal architecture of the S6 block featured in the in architecture show in Figure~\ref{fig:arch}. The first FC layer has a number of units equal to twice the length of the input dimension and, in turn, is used to create a double projection of its input. The projection is split into two equal-sized vectors, one passing the convolutional layer and, afterward, the swish function and, finally, the S6 layer; the other is element-wise multiplied by the output of the S6 layer after the swish function. The result from the residual connection feeds a FC layer with a number of units equal to the length of the block input vector. Next to the layers are the reported number of units (u), activation functions, and kernel (k) when applicable.}\label{fig:Ma}
\end{figure}

The S6 layer is used inside a block, as in ~\citep{gu2023mamba}, illustrated in Figure~\ref{fig:Ma}. The first layer of the block is a linear FC with a number of units equal to double the length of the desired internal dimension, thereby expanding the block's input size. The chosen internal dimension is $3$, corresponding to the output size of the S6 layer, selected to limit the total number of model parameters. The expanded vector is split into two equal-sized vectors, one passing the convolutional layer and, afterward, swish function ~\citep{ramachandran2017searching} and finally the S6 layer; the other is element-wise multiplied by the output of the S6 layer, after the swish function as well. The swish function acts as a gating mechanism to regulate the flow of information and features a learnable parameter, providing smoother behavior than a ReLU. 

Finally, since S6 does not include nonlinear operators, the block output is processed through a nonlinear FC with $2$ units and the Gaussian Error Linear Unit (GELU) ~\citep{hendrycks2023gaussian} activation function. Swish and GELU activation functions, with continuous derivatives, have been preferred because they have shown empirically better performance than ReLU ~\citep{ramachandran2017searching, hendrycks2023gaussian}, a choice also made in ~\citep{gu2022parameterization, gu2023mamba}. Figure ~\ref{fig:Ma} illustrates the described process, including the size of each layer, which can be described as follows:
\begin{align}
[\boldsymbol{s_{p}}, \boldsymbol{s_{res}}] &= FC(\boldsymbol{x_{proj}}) \nonumber \\
\boldsymbol{s_{conv}} &= Conv(\boldsymbol{s_{p}}) \nonumber \\
\boldsymbol{s_{ssm}} &= S6(Swish(\boldsymbol{s_{conv}})) \nonumber \\
\boldsymbol{s_{s}} &= \boldsymbol{s_{ssm}} \otimes Swish(\boldsymbol{s_{res}}) \nonumber \\
\boldsymbol{s_{o}} &= GELU(FC(\boldsymbol{s_{s}})))
\end{align}
The conditioning block processes separately the two distinct groups of control parameters $\boldsymbol{p}$: those influencing the amount of compression $\boldsymbol{p_{co}}$, and those determining the timing behavior $\boldsymbol{p_{ti}}$. Both parameter groups are processed alongside the vector $\boldsymbol{f}$, computed processing the magnitude spectrum of the $64$ input samples $\boldsymbol{x}$ with a convolutional layer. The kernel size of this layer is set to one more than the FFT size used to compute the magnitude spectrum. An FFT size of $128$ was selected based on preliminary experiments aimed at determining the optimal size, with consideration given to limiting the total number of computations.

Compression-related parameters $\boldsymbol{p_{co}}$ concatenated with $\boldsymbol{f}$ derived from the magnitude spectrum are passed to a linear FC layer to compute the modulation vectors $\boldsymbol{a}$ and $\boldsymbol{b}$ to perform the FILM on $\boldsymbol{g_{m}}$:
\begin{equation}
    \boldsymbol{k_{f}} = \boldsymbol{a} \boldsymbol{g_{m}} + \boldsymbol{b}
\end{equation}
The output $\boldsymbol{k_{f}}$ is processed by a GLU layer, where another linear FC layer doubles the input dimensionality, producing $\boldsymbol{k_{f1}}$and $\boldsymbol{k_{f2}}$ which are then used to compute the output $\boldsymbol{k_{g}}$ as follows:
\begin{equation}
    \boldsymbol{k_g} = \boldsymbol{k_{f1}} \odot softsign(\boldsymbol{k_{f2}})
\end{equation}
Timing-related parameters $\boldsymbol{p_{ti}}$, concatenated with $\boldsymbol{f}$, are processed by an additional modulation layer, which in this case utilizes a GRU layer to produce the modulation vectors $\boldsymbol{c}$ and $\boldsymbol{d}$, which are then applied to perform the Temporal FiLM on $\boldsymbol{k_{g}}$:
\begin{equation}
    \boldsymbol{k_{nf}} = \boldsymbol{c} \boldsymbol{k_{g}} + \boldsymbol{d}
\end{equation}
Similar to the previous step, $\boldsymbol{k_{nf}}$ is processed by a GLU layer, followed by another linear FC layer that produces two vectors, $\boldsymbol{k_{f3}}$ and $\boldsymbol{k_{f4}}$. These vectors are then used to compute the output of the conditioning block, denoted as $\boldsymbol{g_c}$, as follows:
\begin{equation}
    \boldsymbol{g_c} = \boldsymbol{k_{f3}} \odot softsign(\boldsymbol{k_{f4}})
\end{equation}

\begin{figure*}[h]%
\centering
\includegraphics[width=0.9\textwidth]{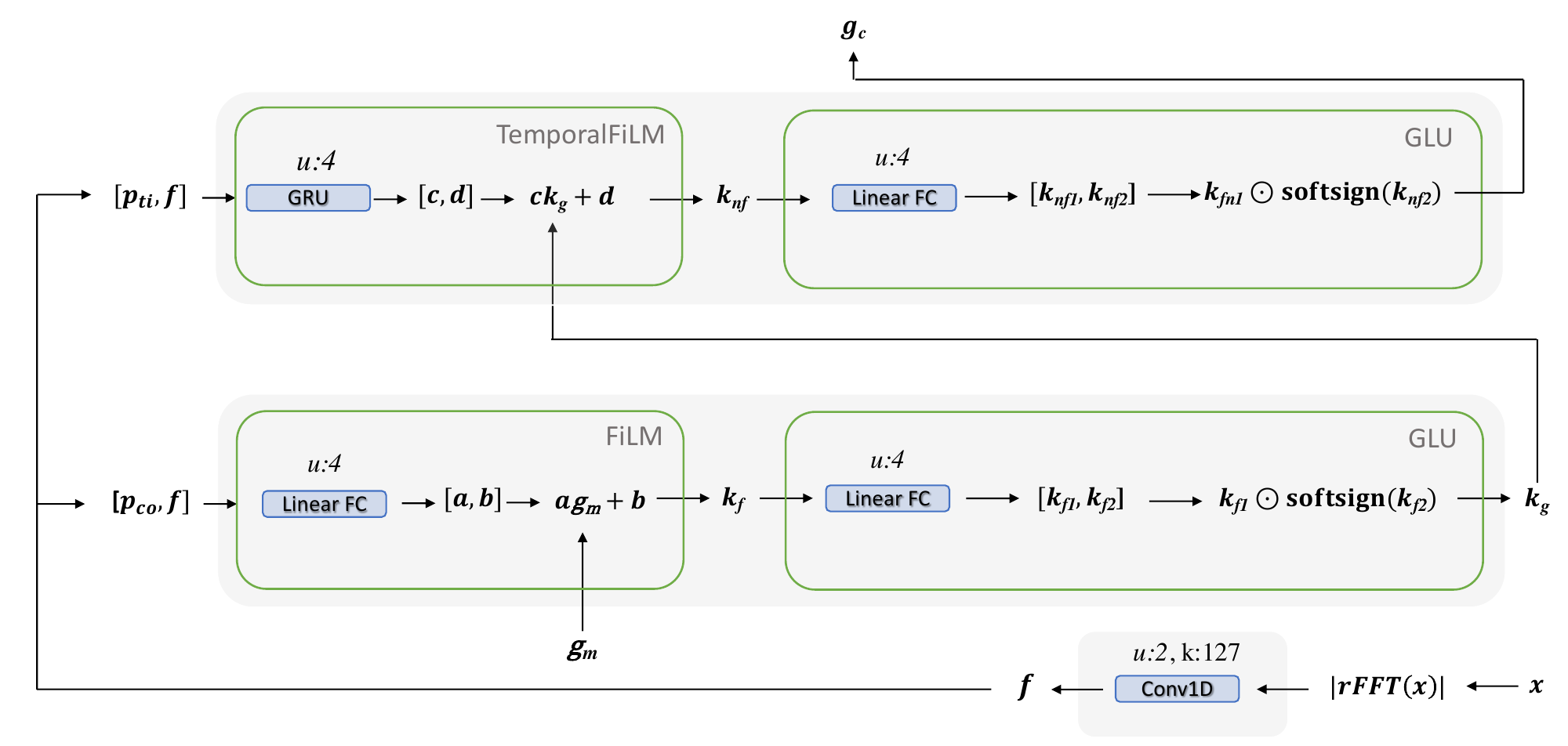}
\caption{Internal architecture of the conditioning block featured in the architecture shown in Figure~\ref{fig:arch}. The device control parameters, conditioning the model, are split into two vectors: those influencing the amount of compression $\boldsymbol{p_{co}}$ and those determining the timing behavior $\boldsymbol{p_{ti}}$. The FiLM method, followed by a GLU featuring the softsign function, is used for $\boldsymbol{p_{co}}$. Similarly, the temporal FiLM method, followed by a GLU featuring the softsign function, is used for $\boldsymbol{p_{ti}}$. In both cases, the vector $\boldsymbol{f}$ computed by a convolutional layer, which takes the magnitude spectrum of the input $\boldsymbol{x}$ is concatenated to $\boldsymbol{p_{co}}$ and $\boldsymbol{p_{ti}}$. In the case of the LA-2A, $\boldsymbol{p_{co}}$ corresponds to the peak reduction and the switch mode, while $\boldsymbol{p_{ti}}$ pertains to the vector $\boldsymbol{f}$. For the CL 1B, $\boldsymbol{p_{co}}$ includes both the threshold and ratio, whereas $\boldsymbol{p_{ti}}$ encompasses the attack and release times. Next to the layers are the reported number of units (u) and kernel (k) when applicable.}\label{fig:cond}
\end{figure*}

The input and output dimensionalities of the conditioning block, as well as those of the internal intermediate quantities, are always equal to $2$. Figure ~\ref{fig:cond} summarizes the conditioning processes based on the device variable control parameters $\boldsymbol{p}$, reporting the number of units and the kernel size, when applicable, for each layer.

As detailed in Section~\ref{sec:datasets}, the LA-2A features a continuous parameter called `peak reduction', which defines the amount of compression, and a switch to toggle between limit and compression modes, which influences the amount of compression of the device. The attack and release times are not controllable and depend on the input signal. Therefore, for the LA-2A, $\boldsymbol{p_{co}}$ includes the `peak reduction' and the binary value of the mode switch, and $\boldsymbol{p_{ti}}$ includes only the vector $\boldsymbol{f}$. On the other hand, the CL 1B offers four controllable parameters: threshold, ratio, attack, and release time. However, the threshold and ratio influence the amount of compression, while the attack and release times affect the timing behavior of the device. Therefore, for the CL 1B, $\boldsymbol{p_{co}}$ includes the values of the threshold and ratio, while $\boldsymbol{p_{ti}}$ includes the values of the attack and release times.

\begin{figure*}[h]%
\centering
\includegraphics[width=0.75\textwidth]{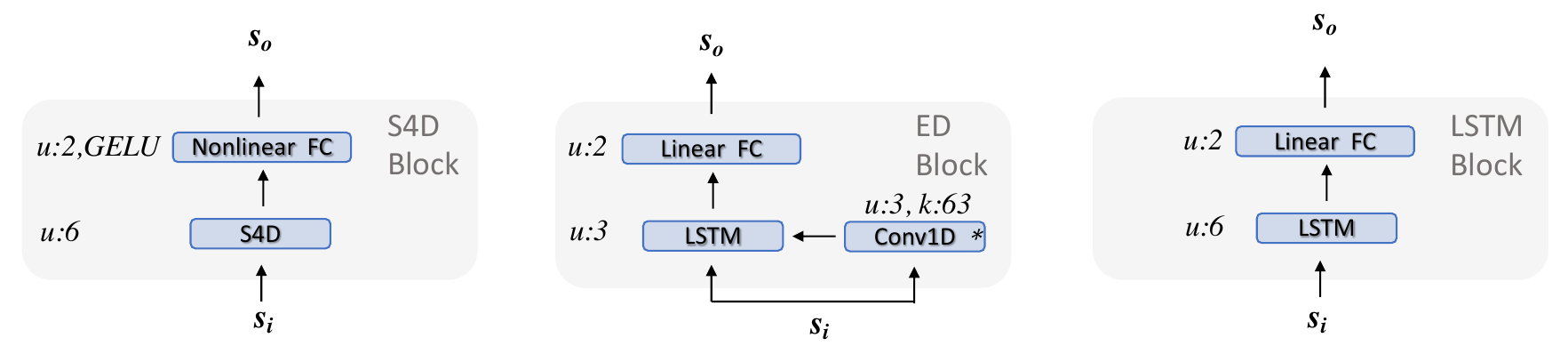}
\caption{Recurrent blocks used as alternative to the S6 block in the Figure~\ref{fig:arch} architecture for comparing performances. From left to right: S4D, ED, and LSTM blocks. In the ED block, the convolutional layer is included only before the conditioning block. Next to the layers are the reported number of units (u), activation function, and kernel (k) when applicable.}\label{fig:all}
\end{figure*}

To evaluate the effectiveness of the proposed architecture in modeling analog optical compression, we designed three additional architectures that share the same conditioning process but feature different recurrent blocks, specifically based on LSTM, ED, and S4D, as illustrated in Figure~\ref{fig:all}. This approach allows us to compare their capabilities in modeling nonlinearities and encoding temporal information and to compare their performance when integrated with the same conditioning process. 

The S4D architecture uses the S4D layer, which is also described by Equation~\ref{eq:ssm}; however, in this case, the $\boldsymbol{B}$, $\boldsymbol{C}$, and $\boldsymbol{D}$ are independent of the input and learned during the training process. As mentioned earlier, the principal difference with the S6 variant lies in this aspect. As the S6, the S4D layer incorporates a FC layer followed by a GELU activation function, as it does not inherently include nonlinearity. The ED architecture utilizes an encoder-decoder structure inspired by our previous work~\citep{simionato2023fully}, which processes the input segment with a convolutional layer to establish the initial internal states for the first LSTM layer. Unlike the first, the second LSTM layer relies on its own internal states. The ED architecture also incorporates a linear FC layer following the LSTM layers. The S4D architecture employs a specialized SSM layer, which is an efficient variant that parameterizes its state matrix as a diagonal matrix ~\citep{gu2022parameterization}. The LSTM architecture consists of an LSTM layer followed by a linear FC layer, embodying the most traditional approach. 
 
In addition to the architectures described above, we also use several previously proposed models as baselines for comparing performance. These models include LSTM-b~\citep{steinmetz2021efficient}, ED-b~\citep{simionato2023fully}, and TCN-b~\citep{steinmetz2021efficient}. Both LSTM-b and ED-b baselines have an identical input, including $64$ samples. They are designed to have a number of trainable parameters comparable to that of the proposed model—with the LSTM-b having $13$ units and the ED-b model having $6$ units for each layer. In contrast, TCNs are subject to specific architectural constraints, such as the number of input and output samples that define the receptive field. For the TCN-b, we opted for the highest-performing model described in the literature without making any modifications. This model processes $7,200$ input samples to generate $57,600$ output samples at each inference iteration—quantities significantly larger than those associated with our proposed architectures.

\subsection{Computational Cost and Latency}\label{sec:Costs}

Our proposed architecture has been designed to achieve minimal computational complexity during inference by limiting the total number of trainable parameters to approximately $1,000$. This constraint has enabled us to reduce the computational load required to process a stereo audio signal at $48$~kHz to just under $200$ MFLOPS (Floating Point Operations Per Second). Such a limit is well-suited for most applications, including running multiple software audio effects concurrently on consumer-grade personal computers. Although implementations on dedicated DSP hardware can generally handle higher loads, we hypothesize that expanding the model size within the same architectural framework could yield more accurate emulations. Furthermore, the deliberate choice of relatively small models has facilitated the conducting of a substantial number of experiments involving architectural, dataset, and training variations, as the number of trainable parameters significantly impacts the training time.

To ensure a fair comparison in our study, we applied the same limit on the number of trainable parameters to the other architectures evaluated. We permitted minor deviations from this limit to accommodate the use of distinct layer types in each case. We have calculated the computational cost of inference for our proposed architecture in terms of Floating Point Operations (FLOPs) per audio sample by starting from the equations defining each layer's operations. Our FLOP calculations include both multiplication and addition operations. For the Softsign, Swish, GELU, and hyperbolic tangent activation functions, we have estimated for $10$ FLOPs.

Table~\ref{tab:costs} presents the computational cost and number of trainable parameters for both the proposed architecture, which includes the S6 block as shown in Figure~\ref{fig:arch}, and for all the variants illustrated in Figure~\ref{fig:all}. Since CL 1B has twice as many control parameters as LA-2A, which are used to condition the model, the proposed architecture yields slightly different computational costs and numbers of trainable parameters when applied to these two distinct analog optical compressor devices. The conditioning block requires $352$ and $400$ FLOPs for CL 1B and LA-2A cases, respectively.

The latency for all models is $64$ audio samples, corresponding to $1.33$~ms at a sampling rate of $48$~kHz, with the exception of the TCN baseline model, which exhibits a latency of $14400$ audio, samples equivalent to $300$~ms. 

\begin{table}[t]
\centering
\tabcolsep8.1pt
\caption{FLOPs for the inference of a single audio sample and the number of trainable parameters for all architectures used in the comparative experiments. Two values are reported for each architecture: the first corresponds to the LA-2A and the second to the CL 1B. The slight differences between the two sets of values are attributable to the varying number of control parameters in the two devices.
}
\label{tab:costs}
{%
\begin{tabular}{@{}llllc@{}}\toprule
 & FLOPs/sample & Parameters\\
\hline
S6 & 1242, 1290 & 984, 1000\\
S4D & 1220, 1268 & 1027, 1043 \\
ED & 1252, 1300 & 1009, 1025\\ 
LSTM & 1668, 1716 & 989, 1005\\ 
\hline
ED-b & 1762, 1794 & 1005, 1021\\ 
LSTM-b & 2453, 2869 & 994, 994 \\
TCN-b & 467 & 8795185, 8795217\\ 
\hline%
\end{tabular}}
\end{table}

\section{Methods}\label{sec:methods}

We have conducted an extensive set of experiments to assess the modeling accuracy of the proposed model, which is based on a S6 architecture, against state-of-the-art alternatives, including LSTM, ED, S4D, and TCN architectures. Evaluations are based on quantitative metrics and listening tests. We have collected and used different datasets to evaluate the models. Specifically, in addition to data gathered from physical analog compressors, we have also collected data from their commercial software emulations to assess how dataset size and organization affect modeling accuracy.

\subsection{Datasets}\label{sec:datasets}

\begin{figure}[h]%
\centering
\includegraphics[width=0.47\textwidth]{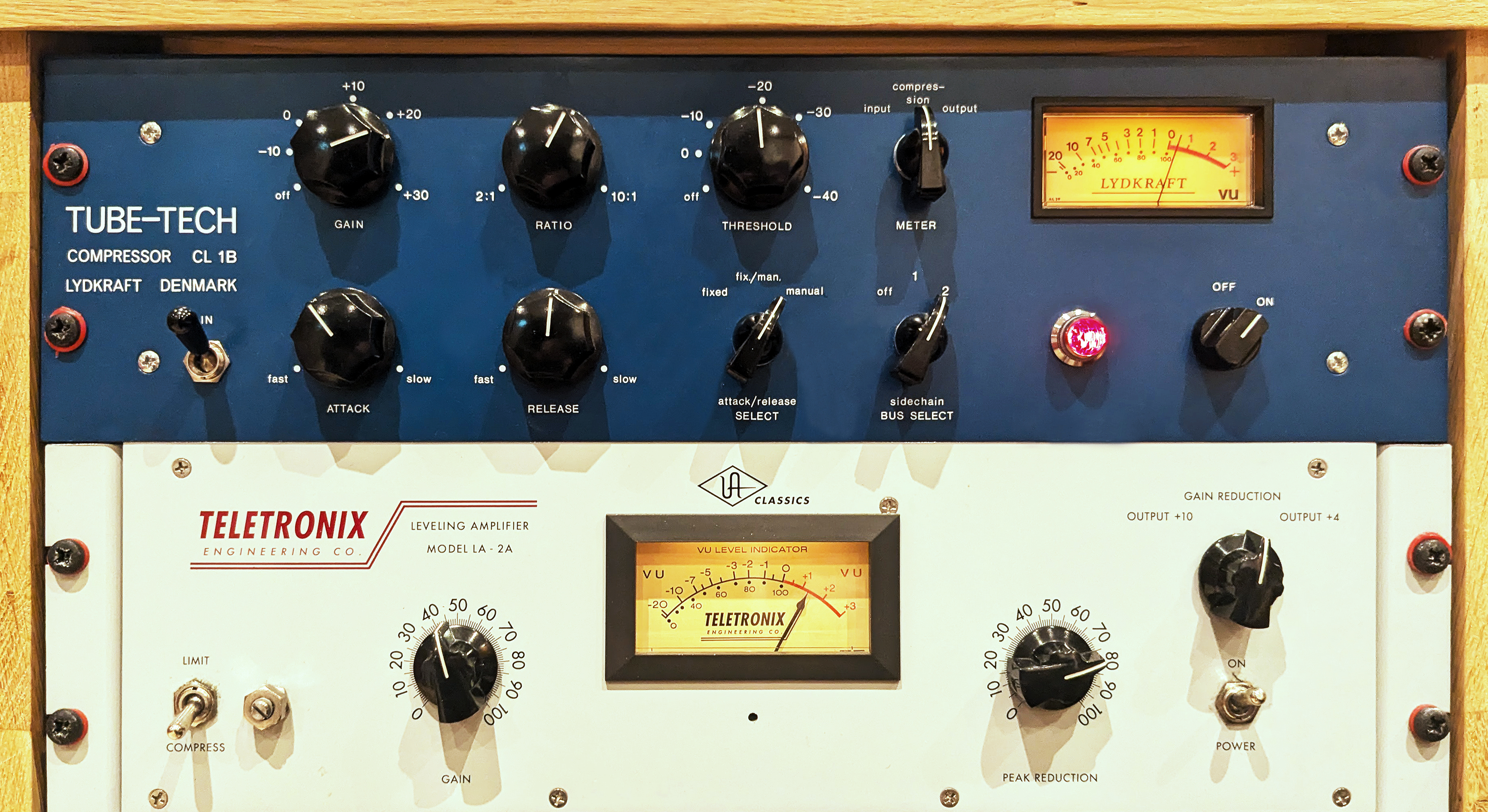}
\caption{TubeTech CL 1B (top) and Teletronix LA-2A (bottom) optical range dynamic compressors used to collect the dataset for the experiments.}\label{fig:comp}
\end{figure}

The datasets include recordings from both analog hardware devices and their digital software emulations. Specifically it includes recordings from the TubeTech CL 1B~\footnote{\url{http://www.tube-tech.com/cl-1b-opto-compressor/}} and Teletronix LA2A~\footnote{\url{https://www.uaudio.com/hardware/la-2a.html}}, shown in Figure~\ref{fig:comp}, along with their software emulations: Softube CL 1B~\footnote{\url{https://www.softube.com/tube-tech-cl-1b-mk-ii}} and the Universal Audio LA-2A~\footnote{\url{https://www.uaudio.com/uad-plugins/compressors-limiters/teletronix-la-2a-tube-compressor.html}}.

The CL1B, manufactured by TubeTech, is an analog optical compressor featuring five variable control parameters: ratio, threshold, attack, release
times, and output gain. The threshold can be set from $0$
to $-40$ dBu, the ratio from 2:1 to 10:1, the attack time from $0.5$ to $300$~ms, and release time from $0.05$ to $10$~seconds. Since the precise values for attack and release times are not marked around the associated knobs of the device, we have assumed a linear relationship between the minimum and maximum values, as indicated in the manual. The device features an output gain with vacuum tube-based circuitry that can produce harmonic distortion, two sidechain buses, and three time-constant circuit modes: fixed, fixed/manual, and manual. The first mode offers fixed attack and release times ($1$~ms and $50$~ms, respectively); the second mode provides a fixed attack time of $1$~ms with variable release; and the third mode allows for variable attack and release times.

The LA-2A Leveling Amplifier, an analog limiter/compressor manufactured by Teletronix, shares the optical gain reduction feature with the CL 1B but has distinct operational characteristics. Unlike the CL 1B, the LA-2A's attack and release times are not adjustable; it has a fixed average attack time of $10$  ms and a two-stage release. The first stage lasts $0.06$ seconds, followed by a variable second stage determined by the light-sensitive resistor's response, which can extend from $0.5$ to $5$~seconds. This variability means the LA-2A's release time is signal-dependent and can lengthen if the compression is heavy or the signal consistently exceeds the threshold. Consequently, the release time for the LA-2A cannot be known a priori, as it is highly dependent on the signal's history. The device features an input gain control knob, a switch to toggle between limiter or compressor mode, and a peak reduction knob that adjusts the amount of compression. Both knobs have markings with values that range from $0$ and $100$.

To collect the data, we used a MOTU M4 audio interface to send a mono audio signal to the compressor's input and simultaneously record its output. Additionally, we connected one output channel of the audio interface directly to its input, allowing us to simultaneously and synchronously record both the input signal being sent to the compressor and the processed output signal. For training the models, we used the input recorded via audio interface loopback, compensating for the interface's minor sound coloring and latency. The input mono signal, with a duration of $195$ seconds, includes frequency sweeps covering a range of $20$ Hz to $20$ kHz, white noises with increasing amplitudes (both linear and logarithmic), recordings of instruments such as guitar, bass, drums (individual notes and riffs), vocals, piano, pad sounds, and sections of electronic and rock songs.

We recorded different pairs of input and output signals, using a sampling rate of $48$~kHz, for various settings of the compressor's control parameters. For the Teletronix LA-2A, we captured the compressor's output for $10$ different values of the peak reduction parameter, evenly spaced between the minimum and maximum settings. This process was repeated with the device set to limiter and compressor modes, resulting in $20$ unique parameter combinations and their corresponding input-output signal pairs. The input gain was set to $23$, a setting we empirically determined to match the input and output levels with no compression applied. For the TubeTech CL 1B, we selected $3$ different values per parameter, evenly spaced between the minimum and maximum settings, resulting in a total of $81$ unique parameter combinations for which we recorded the input and output signals. When collecting the dataset, the output gain was set to $0$, the sidechain bus was turned off, and the time-constant circuit mode was turned to manual.

For the software emulations, we created the dataset using a fully automated software tool that we specifically developed to generate datasets from audio effects ~\citep{fasciani2024auniversal}. This tool enables the easy generation of datasets with a much larger and more precisely controlled combination of parameters than is possible with hardware analog devices, where the parameters are not controllable via an electrical interface. The dataset from the Universal Audio LA-2A software emulation was generated considering $100$ equally spaced (EQS) values that cover the full range of the peak reduction in both limiter and compressor modes. Also, in this case, the input gain was set to $23$. 

For the Softube CL 1B emulation, which has four variable parameters, increasing the number of values selected for each parameter significantly raises the number of unique parameter combinations and, consequently, the size of the dataset. This increment may lead to severe computational challenges when training the models, potentially rendering the training process infeasible. Therefore, for the Softube CL 1B, we generated two separate datasets: the first, tagged as `dynamic', includes a variable ratio and threshold with attack and release times knobs fixed at $25\%$, which should correspond to $70$~ms and $1$~second, respectively; the second, tagged as `time', includes variable attack and release times with the ratio knob fixed $50\%$, which should correspond to $6:1$, and the threshold at $-20$~dBu. In both Softube CL 1B datasets, the variable parameters consist of $20$ EQS values that cover the full range of each parameter. Also, in this case, the output gain was set to $0$, the sidechain bus was turned off, and the time-constant circuit mode was turned to manual.

A summary of the datasets, including their size, duration, and parameter values, is provided in Table~\ref{tab:summary}. Datasets are publicly available online~\citep{riccardo_simionato_2024}.

\begin{table*}[t]
    \begin{center}
	\begin{tabular}{|l|l|l|l|l|l|}
		\hline
		Dataset & Combinations & Duration & Parameters & Range & Values \\
		\hline
		\texttt{Teletronix LA-2A} & 20 & 3,900 s & Peak Reduction & [0, 100] & 10 EQS \\
    	& & & Switch & [0, 1] & [0, 1] \\
     \hline
            \texttt{TubeTech CL 1B}  & 81 & 15,795 s &  Ratio & [2:1, 10:1] & 3 EQS\\
    	& &  & Threshold & [0, -40] dBu  & 3 EQS\\
            & &  & Attack & [0.5, 500] ms  & 3 EQS\\
            & &  & Release & [0.005, 10] s  & 3 EQS\\
             \hline
             \hline
    \texttt{Universal Audio LA-2A} & 200 & 39,000 s & Peak Reduction  & [0, 100] & 100 EQS\\
    	& &  & Switch & [0, 1] & [0, 1]\\ 
     \hline
            \texttt{Softube CL  1B}  & 400 & 78,000 s &  Ratio & [2:1, 10:1] & 20 EQS\\
    	\texttt{dynamic} & & & Threshold & [0, -40] dBu  & 20 EQS\\
            & &  & Attack & $\sim 70$ ms  & fixed \\
            & &  & Release & $\sim 1$ s  & fixed \\
            \hline
            \texttt{Softube CL 1B}  & 400 & 78,000 s &  Ratio & $\sim 6:1$ & fixed \\
    	\texttt{time} & & & Threshold & -20 dB  & fixed \\
            & &  & Attack & [0.5, 500] ms & 20 EQS\\
            & &  & Release & [0.005, 10] s  & 20 EQS\\
		\hline
	\end{tabular}
    \end{center}
\caption{Overview of datasets collected from the analog compressors and their software emulations, indicating dataset name, parameter combinations, total duration, and parameter sampling details.}
\label{tab:summary}
\end{table*}

\subsection{Training}

All models included in the experiments are trained using the Mean Squared Error (MSE) as a loss function, which has been demonstrated to effectively capture the dynamic aspects of the output signal, as shown in~\citep{simionato2023fully}. The trainings are conducted for a maximum of $200$ epochs, with an early stopping condition triggered if no reduction in validation loss is observed for $30$ consecutive epochs. Models are trained using the input-output recordings in the dataset at the original sampling rate of $48$~kHz. The batch size is set to $2400$ examples, each consisting of $64$ input audio samples, the values of the control parameters, and the single output audio sample. 

Considering the network architecture described in Section~\ref{sec:nn}, in which a buffer of the $64$ most recent samples is used as input, training layers with states necessitates an internal sequential order among the input-output examples. Consequently, our implementation adopts an input shape of ($B$, $L$, $F$), which is set to ($B$, $1$, $64$), where $B$ denotes the arbitrary batch size, $L$ the sequence length, and $F$ the input dimensionality, often identified as the number of input features. This configuration considers the $64$ input samples as the vector of input features at the current sampling instant. This approach aids the network in making predictions while maintaining a stateful design using truncated backpropagation through time. In contrast, a configuration with a sequence length of $64$ leads to more computations for the state updates, exacerbating the vanishing gradient problem and increasing the computational cost and training time of the network. 

The control parameter values are scaled to the range [$0$, $1$], corresponding to the full range of the device. We use the Adam \citep{kingma2014adam} optimizer with a gradient norm scaling of $1$ ~\citep{pascanu2013difficulty}, and we designed an exponential decay learning rate ($lr$) as follows:
\begin{equation}\label{eq:mfcc}
    lr = 0.25^e * LR, \text{    with    } LR = 3 \cdot 10^{-4}
\end{equation}
where LR is the initial learning rate, and $e$ is the number of epochs. The policy for splitting the dataset between training and test sets varies depending on the specific aspect we aim to evaluate, as detailed in Section~\ref{sec:experiments}. Test losses and evaluation metrics are computed using the model’s weights, which minimize the validation loss throughout the training epochs.

\subsection{Experiments}\label{sec:experiments}

The proposed model, which incorporates S6 layers into its architecture, is evaluated against various alternative and baseline architectures, including those featuring LSTM, ED, S4D, and TCN layers. These models are trained on a dataset obtained from the analog optical dynamic range compressors Teletronix LA-2A and TubeTech CL 1B. Trainings use $90\%$ of the dataset, while the remaining $10\%$ is reserved for testing. The $90-10\%$ split is implemented at the individual recording level. Each recording represents the response for a unique combination of control parameters, which remain fixed during the recording process. This splitting strategy ensures that all combinations of control parameters are equally present in both the training and test sets, with the aim of developing an accurate model that replicates the behavior of the actual analog device. 

The initial $90\%$ of each recording is used for the train set, and the remaining portion is allocated to the test set. Manual adjustments are occasionally made to ensure that the splitting points occur during brief intervals of silence between sound sequences in the dataset. Modeling accuracy is assessed using the test set loss and additional quantitative metrics outlined in Section~\ref{sec:metrics}, and through listening tests (excluding the baseline models), as explained in Section~\ref{sec:listening}.

In previously described experiments, models were evaluated on their ability to compress audio signals unseen during training but for control parameter combinations encountered during training. We conducted additional experiments to assess the model's capability to compress signals for control parameter combinations not seen during training. We also evaluated the impact of dataset size, specifically regarding the density of sampled values within the range of each control parameter. For these evaluations, we utilized the three datasets collected from the software emulations Universal Audio LA-2A and Softube CL 1B. Their larger sizes and higher parameter sampling densities allow for experimentation with various subsets of the datasets. The testing framework utilized three datasets: Universal Audio LA-2A, Softube CL 1B `dynamic', and Softube CL 1B `time'. Each dataset is used three times with a fixed test set, while the size of the training is varied. This results in a total of nine comparative experiments. The test set comprises recordings, each with a duration of $195$ seconds, associated with combinations of control parameters that have normalized values of $0.25$, $0.55$, or $0.75$ (within the unit range).

For the Universal Audio LA-2A dataset, this results in a total of $6$ combinations, with $3$ levels of peak reduction for each mode switch position. For the Softube CL 1B `dynamic' and `time' datasets, this yields $9$ combinations for each dataset, reflecting the two variable control parameters. The remaining portions of these datasets are used to create three separate training sets: the first includes all remaining recordings; for the second and third, we progressively reduce the number of EQS values selected for each variable parameter. For the Universal Audio LA-2A dataset, the reduction is from $100$ down to $20$ and $10$. For the Softube CL 1B `dynamic' and `time' datasets, the reduction goes from $20$ down to $10$ and $5$. A summary of these training and test sets is provided in Table~\ref{tab:summarytraintest}.

To further assess how well the models capture the time-dependent response of the compressors and learn to emulate the time-behavior variations caused by the devices' control parameters, we conducted an additional set of experiments. These experiments exclusively utilized TubeTech CL 1B recordings because this device allows for explicit compression attack and release times adjustment. In contrast, the LA 2A's release time is influenced by the input signal. We selected four compression cases from the TubeTech CL 1B dataset, with the threshold set to $-10$~dBu and the ratio fixed at $6:1$. These settings result in discernible compression that is clearly noticeable but not extreme. For each case, the attack and release times differ, spanning from the fastest attack time of $0.5$~ms to the slowest of $500$~ms and from the quickest release time of $0.005$~seconds to the longest of $5$~seconds. We excluded the dataset's slowest release setting of $10$~seconds since it is less commonly used in practical applications and is less suitable for assessing accuracy with short sound samples, as we do in this experiment. 

In the four selected settings, CL 1B presents a very diverse time-dependent response, as the gain reduction is applied and removed with significantly different temporal patterns. We train all models, including the S6, alternative, and baseline architectures, on an exclusive subset of the dataset corresponding to one of the four selected compression settings. This approach aims to determine how well each model can learn specific temporal compression patterns and identify whether these diverse cases pose greater challenges. For testing, we utilize a short, $3.60$-second audio snippet that is not used for training and features bass and drum sounds. This compressed snippet has pronounced differences with the four distinct settings, and any modeling inaccuracies in compression attack and/or release are evident through the loss function and quantitative evaluation metrics. 

Finally, we train each model using all four compression settings to gauge the models' ability to handle multiple parametric time-dependent responses. This allows us to observe any degradation in the metrics that occurs when the model is conditioned with the selected compression settings. Given the small size of the training dataset and quicker convergence compared to previous experiments, we repeated each training 20 times. We report and analyze the performance metrics based on the average of these training iterations.

Experiments were conducted on workstations equipped with an Intel Core i9-14900KF CPU, 192GB of DDR5 RAM at 5200MHz, and a GeForce RTX 4090 GPU. The models were implemented in Python using Tensorflow~\footnote{https://www.tensorflow.org/}. Training times varied across datasets and models, which are summarized in Table~\ref{tab:times}.

\begin{table*}[t]
    \begin{center}
	\begin{tabular}{|l|l|l|l|l|}
	\hline
Set & Combinations & Duration & Parameters & Values  \\
\hline
\texttt{Universal Audio LA 2A test} & 6 & 1,170 s & Peak Reduction  & [0.25, 0.55, 0.75]\\
& & & Switch & [0, 1]\\ 
\hline
\texttt{Universal Audio LA 2A train-0.01} & 194 & 37,830& Peak Reduction  & 100 EQS\\
& & & Switch & [0, 1]\\ 
\hline
\texttt{Universal Audio LA 2A train-0.05} & 34 & 6,630 s& Peak Reduction  & 20 EQS\\
& & &Switch & [0, 1]\\ 
\hline
\texttt{Universal Audio LA 2A train-0.10} & 20 & 3,900 s & Peak Reduction  & 10 EQS\\
& & &Switch & [0, 1]\\ 
\hline
\hline
\texttt{Softube CL 1B dynamic test} & 9 & 1,755 s & Ratio & [0.25, 0.55, 0.75]\\
& & & Threshold & [0.25, 0.55, 0.75]\\
\hline
\texttt{Softube CL 1B dynamic train-0.05} & 391 & 76,245 s & Ratio & 20 EQS\\
& & & Threshold & 20 EQS\\
\hline
\texttt{Softube CL 1B dynamic train-0.10} & 100 & 19,500 s & Ratio & 10 EQS\\
& & & Threshold & 10 EQS\\
\hline
\texttt{Softube CL 1B dynamic train-0.20} & 25 & 4,875 s & Ratio & 5 EQS\\
& & & Threshold & 5 EQS\\
\hline
\hline
\texttt{Softube-CL1B-time-test} & 9 & 1,755 s & Attack & [0.25, 0.55, 0.75]\\
& & & Release & [0.25, 0.55, 0.75] \\
\hline
\texttt{Softube CL 1B time train-0.05} & 391 &  76,245 s & Attack & 20 EQS\\
& & & Release & 20 EQS\\
\hline
\texttt{Softube CL 1B time train-0.10} & 100 & 19,500 s & Attack & 10 EQS\\
& & & Release & 10 EQS\\
\hline
\texttt{Softube CL 1B time train-0.20} & 25 &  4,875 s & Attack & 5 EQS\\
& & & Release & 5 EQS\\
\hline
\end{tabular}
\end{center}
\caption{Overview of the training and test set splits for the Universal Audio LA-2A and Softube CL 1B datasets, indicating the parameter combinations, total duration, and details of the variable parameters with values normalized to the unit range. Fixed parameter values are provided in Table ~\ref{tab:summary}. The numerical value at the end of the name of the training set indicates the step at which the normalized values are sampled. }
\label{tab:summarytraintest}
\end{table*}
\begin{table*}[t]
    \begin{center}
	\begin{tabular}{|l|l|l|l|l|l|l|l|}
	\hline
     & \multicolumn{7}{|c|}{Average Training Time per Epoch (min)}\\
\cline{2-8}
Set &   S6 & S4D & ED & LSTM  & ED-b &  LSTM-b & TCN-b \\
\hline
\texttt{Teletronix LA-2A} & $7$ & $7$ & $7 $ & $12 $ & $4 $ & $4 $ & $1$ \\
\texttt{TubeTech CL 1B} & $37$ & $ 35$ & $ 34$ & $ 35$ & $ 23$ & $ 22$ & $ 3$\\
\texttt{Universal Audio LA-2A-0.01} & $ 25$ & $ 23$ & 
$ 22$ & $ 36$ & $ 24$ &  $ 24$ & $1$ \\
\texttt{Universal Audio LA-2A-0.05} & $ 5 $ & $ 5 $ & $ 5 $ & $ 5 $ & $ 6 $ & $ 5 $ & $1$\\
\texttt{Universal Audio LA-2A-0.10} & $ 2 $ & $ 5 $ & $ 4 $ & $ 4 $ & $ 2 $ & $ 2 $ & $1$\\
\texttt{Softube CL 1B dynamic-0.05} & $ 51 $ & $ 51 $ & $ 59 $ & $ 54 $ & $ 48 $ & $ 45 $ & $ 5 $\\
\texttt{Softube CL 1B dynamic-0.10} & $ 27 $ & $ 27 $ & $ 26 $ & $ 25 $ & $ 17 $ & $ 16 $ & $1 $\\
\texttt{Softube CL 1B dynamic-0.20} & $ 7 $ & $ 7 $ & $ 15 $ & $ 15 $ & $ 11 $ & $ 10 $ & $ 1 $\\
 \texttt{Softube CL 1B time-0.05}  & $ 51 $ & $ 51 $ & $ 59 $ & $ 54 $ & $ 48 $ & $ 45 $ & $ 5 $\\
\texttt{Softube CL 1B time-0.10}  & $ 26 $ & $ 27 $ & $ 25 $ & $ 25 $ & $ 17 $ & $ 18 $ & $1$\\
 \texttt{Softube CL 1B time-0.20}  & $ 7 $ & $ 7 $ & $ 14 $ & $ 14 $ & $ 12 $ & $ 12 $ & $ 1 $\\
\hline
\end{tabular}
\end{center}
\caption{Average training time per epoch, including validation, across different datasets. Times are expressed in minutes.}
\label{tab:times}
\end{table*}

\subsubsection{Metrics}\label{sec:metrics}

We use various metrics to evaluate the model's accuracy from different perspectives. These metrics include the Mean Absolute Error (MAE) and Error-to-Signal Ratio (ESR), Root-Mean-Square Error (RMSE), Spectral Flux Error (SFE), Multi-resolution Short-Time Fourier Transform Error (M-STFTE), and the Perceptual Similarity Measure (PSM).

The RMSE is defined as:
\begin{align}
RMSE = \sqrt{\frac{1}{N}\sum_{n=1}^{N} (|y_n|^2 - |\hat{y}_n|^2)}
\end{align}
where $y_n$ and $\hat{y}_n$ represent the target and predicted signal, respectively, and $N$ is the length, in samples, of the signals that are considered for computing the metric. The RMSE provides insights into the energy deviation between the target and the prediction.

We employ the SFE to address prediction errors occurring during signal transients. Other losses and metrics calculate the average difference between the target and predicted signal, which can hide significant errors that occur only during sharp attacks or decay and may represent a small fraction of the overall duration of the test signal. The SF error considers the differences between pairs of spectra computed on consecutive overlapping windows and is defined as:

\begin{equation}\label{eq:diff}
\begin{split}
  SFE = \Big|\Big|  \Big( \Big| |STFT(y_n)| - |STFT(y_{n-1})| \Big| \Big)   \\ 
    - \Big(\Big| |STFT(\hat{y}_n)| - |STFT(\hat{y}_{n-1})| \Big| \Big) \Big|\Big|_1
    \end{split}
\end{equation}
where the STFT is computed with windows of $2048$ samples and $75\%$ overlap. This metric emphasizes the ability to predict rapid changes in the output signal accurately.

To assess the accuracy of the model's frequency response, we use the M-STFTE error, as described in~\citep{engel2020ddsp}, normalized with the magnitude of the target. This metric compares the linear spectral distances with varying frequency resolutions using the L1 norm and is defined as:
\begin{equation}\label{eq:stft}
\begin{split}
    M\textrm{--}STFTE &= \frac{1}{N} \sum_m \frac{|| |STFT_m (y_n)| - |STFT_m (\hat{y}_n)| ||_1}{||STFT_m (y_n)||} \\
    \end{split}
\end{equation}
The $STFT_m$ are computed windows of [$512$, $1024$, $2048$] samples $75\%$ overlap. The STFT error is calculated as the sum of the absolute values of the difference vectors. This metric quantifies the error between the spectra at different resolutions.

Lastly, we used the PSM derived from the PEMO-Q auditory model ~\citep{huber2006pemo}. This metric calculates the correlation between internal ear representations as estimated by psychoacoustic models. The PSM measures the perceived audio quality of any audio signal and its distortions, yielding a value within the range of [$-1, 1$]. 

\subsubsection{Listening Tests}\label{sec:listening}

We use Mushra tests~\citep{bs20151534} to evaluate the perceptual similarity of the compressed signal produced by different models against the actual reference recording taken from the analog optical compressors. Listening tests are only carried out for models trained on the TubeTech CL 1B and Teletronix LA-2A datasets. The tests are conducted using the WebMUSHRA platform ~\citep{schoeffler2018webmushra} and involve only participants with music performance or production expertise. For the test, we utilize a standardized setup consisting of a laptop, studio-grade audio interface, and high-end studio headphones.

Initially, the participants are briefed on the test's objectives and given instructions on using the listening test's graphical user interface. Each session comprised six different listening tests conducted in a fixed order. Participants are asked to rate the following eight signals on a scale from $0$ to $100$ in comparison to the reference recording: the original `hidden' reference signal, the reference signal with low-pass filtering at $7$~kHz and $3.5$~kHz, the outputs of the proposed model with the S6 block, alternative models with recurrent blocks based on S4D, ED, and LSTM, and the output of commercial software emulation of the specific compressor, denoted as SW. The test signals, lasting 5 seconds each, covered a variety of drums, guitar, and piano sounds. Participants are allowed to listen to the audio samples repeatedly to ensure accurate evaluation. 

Of the six tests, three pertained to models of the Teletronix LA-2A compressor and three to the TubeTech CL 1B. Baseline models were excluded after pilot listening tests as their ratings were significantly lower than those of other models, a finding expected by informal listening and the higher loss values listed in Table~\ref{tab:summaryanalog}. The initial inclusion of baseline models also led to longer test durations and detrimental listening fatigue among participants since the number of stimuli approached the recommended maximum ~\citep{bs20151534}.

The examples included in the listening test feature different settings for the compression and timing parameters taken from the test set of the respective dataset. These represent typical usage scenarios, excluding extreme compression settings—both heavy, which are infrequently used, and subtle, which can be difficult to detect, even for professional audio engineers. Results from participants who rated the hidden reference lower than $90$ or the $3.5$ kHz low-passed anchor higher than $90$ in $15\%$ or more of the test are discarded. The results were summarized using minimum and maximum values, median, and lower and upper quartiles.

\section{RESULTS AND ANALYSIS}\label{sec:results}

Table~\ref{tab:summaryanalog} reports the results for models trained on the Teletronix LA-2A and TubeTech CL 1B datasets, detailing loss and quantitative evaluation metrics based on test set performance. The table clearly shows that the proposed model with S6 layers outperforms all others in replicating analog optical compressors' time-dependent nonlinear sound alteration processes—processes that vary according to each device's specific control parameters. This is particularly true for the TubeTech CL 1B, where the performance gap between the S6 and all the other models is notably larger. 

In contrast, for the Teletronix LA-2A, the S6 model performs more closely to the S4D, ED, and LSTM models. In both cases, the proposed S6 model achieves lower errors for all considered metrics, with the exception of PSM, which is less indicative due to values close to $1$ for all the models. This suggests minimal differences among the models according to the PEMO-Q auditory perception model and underscores the need for actual listening tests. While S4D and ED models exhibit similar behavior on this modeling task, the LSTM models produced the worst results. A similar trend is observed with LSTM-b among the baselines. Lastly, ED-b demonstrates notably high accuracy for the Teletronix LA-2A dataset.
\begin{table*}[t]
\centering
\tabcolsep8.1pt
\caption{Loss and performance metrics for models trained on the \texttt{TubeTech CL 1B} and \texttt{Teletronix LA-2A} datasets, measured on the test set.}
\label{tab:summaryanalog}
{%
\small
\begin{tabular}{@{}llllllllc@{}}\toprule
Dataset & Model & MSE & MAE & ESR & RMSE & SFE & M-STFTE & PSM\\
\hline
LA2A   & \textbf{S6} & $ 2.39 \cdot 10^{-5}$ & $ 2.32 \cdot 10^{-3}$ & $ 2.23 \cdot 10^{-2}$ & $ 2.23 \cdot 10^{-3}$  & $4.59\cdot 10^{-3}$ & $0.46$ & $0.98$\\
&  S4D & $ 2.57 \cdot 10^{-5}$ & $ 2.42 \cdot 10^{-3}$ & $ 2.40 \cdot 10^{-2}$ & $ 2.33 \cdot 10^{-3}$  & $5.16\cdot 10^{-3}$ & $0.54$ & $0.98$\\
& ED & $ 3.23 \cdot 10^{-5} $ & $ 2.87\cdot 10^{-3}$ & $ 3.01 \cdot 10^{-2}$ & $2.78 \cdot 10^{-3}$ & $5.60\cdot 10^{-3}$ & $0.76$ & $0.99$\\
& LSTM & $ 4.82\cdot 10^{-5}$ & $ 3.42\cdot 10^{-3}$ & $ 4.50\cdot 10^{-2}$ & $ 3.34\cdot 10^{-3}$ & $5.38\cdot 10^{-3}$ & $0.73$ & $0.99$\\
\hline
& ED-b & $8.15 \cdot 10^{-5}$ & $5.46 \cdot 10^{-3}$ & $7.61 \cdot 10^{-2}$ & $5.12 \cdot 10^{-3}$ & $7.22\cdot 10^{-3}$ & $0.78$  & $0.98$\\
 & LSTM-b & $8.06\cdot 10^{-4}$ & $ 1.36 \cdot 10^{-2}$ & $ 7.52\cdot 10^{-1}$ & $ 1.36\cdot 10^{-2}$ & $1.02\cdot 10^{-2}$ & $2.93$ & $0.97$\\
& TCN-b & $3.59 \cdot 10^{-4}$ & $1.30 \cdot 10^{-2}$ & $3.39 \cdot 10^{-1}$ & $1.16 \cdot 10^{-2}$ & $1.87\cdot 10^{-2}$& $1.43$ & $0.97$\\
\hline
\hline
CL1B &\textbf{S6} & $5.86\cdot 10^{-5} $ & $3.69\cdot 10^{-3}$ & $9.61\cdot 10^{-2}$ & $3.67 \cdot 10^{-3}$ & $ 1.03\cdot 10^{-2}$ &$1.46$ & $0.94$\\
& S4D & $ 1.95 \cdot 10^{-4}$ & $6.86 \cdot 10^{-3}$ & $ 3.19 \cdot 10^{-1}$ & $ 6.75 \cdot 10^{-3}$ & $1.04\cdot 10^{-2}$ & $2.20$ & $0.95$\\
& ED & $ 1.19 \cdot 10^{-4}$ & $ 5.67 \cdot 10^{-3}$ & $ 1.95 \cdot 10^{-1}$ & $ 5.63 \cdot 10^{-3}$ & $1.67\cdot 10^{-2}$& $1.98$  & $0.92$\\
& LSTM & $2.96 \cdot 10^{-4}$ & $ 8.53  \cdot 10^{-3}$ & $ 4.85  \cdot 10^{-1}$ & $ 8.43 \cdot 10^{-3}$ & $1.52\cdot 10^{-2}$ & $2.65$ & $0.94$\\
\hline
& ED-b & $ 4.31\cdot 10^{-4}$ & $ 1.14 \cdot 10^{-2}$ & $  7.07\cdot 10^{-1}$ & $1.01 \cdot 10^{-2}$&  $2.13\cdot 10^{-2}$ & $2.80$ & $0.90$\\
& LSTM-b & $ 4.88 \cdot 10^{-4}$ & $1.16 \cdot 10^{-2}$ & $ 8.01 \cdot 10^{-1}$ & $ 1.07\cdot 10^{-2}$ & $2.89\cdot 10^{-2}$ & $2.91$ & $0.90$\\
& TCN-b & $4.05 \cdot 10^{-4}$ & $1.21 \cdot 10^{-2}$ & $ 6.74\cdot 10^{-1}$ & $1.08 \cdot 10^{-2}$& $1.41\cdot 10^{-2}$& $2.74$ & $0.95$\\\hline
\end{tabular}}
\end{table*}

\begin{figure}[h]%
\centering
\includegraphics[width=0.45\textwidth]{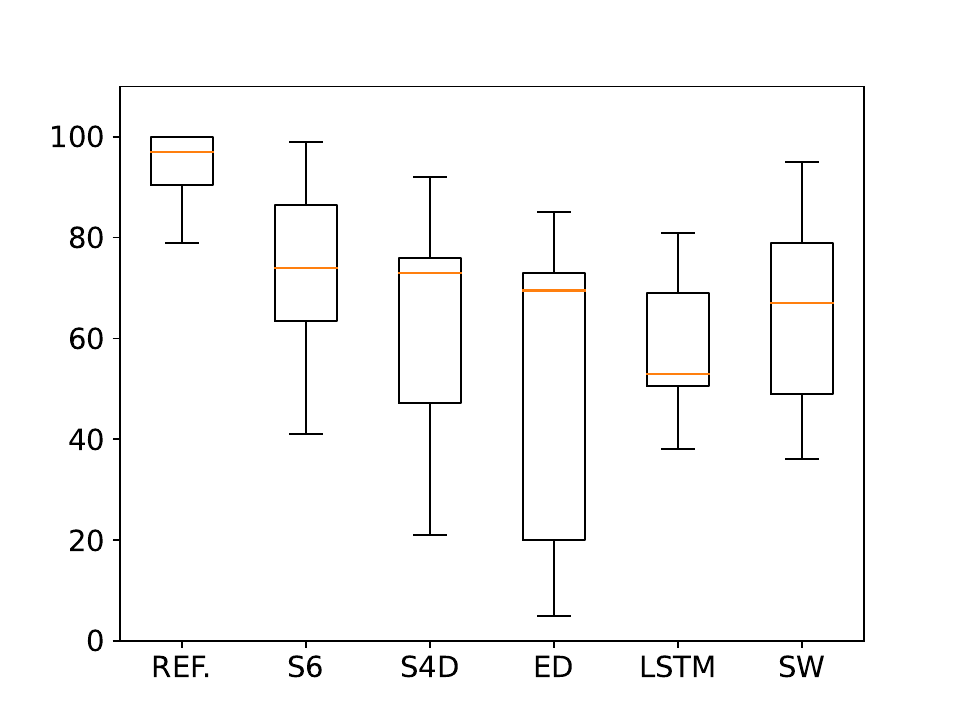}
\includegraphics[width=0.45\textwidth]{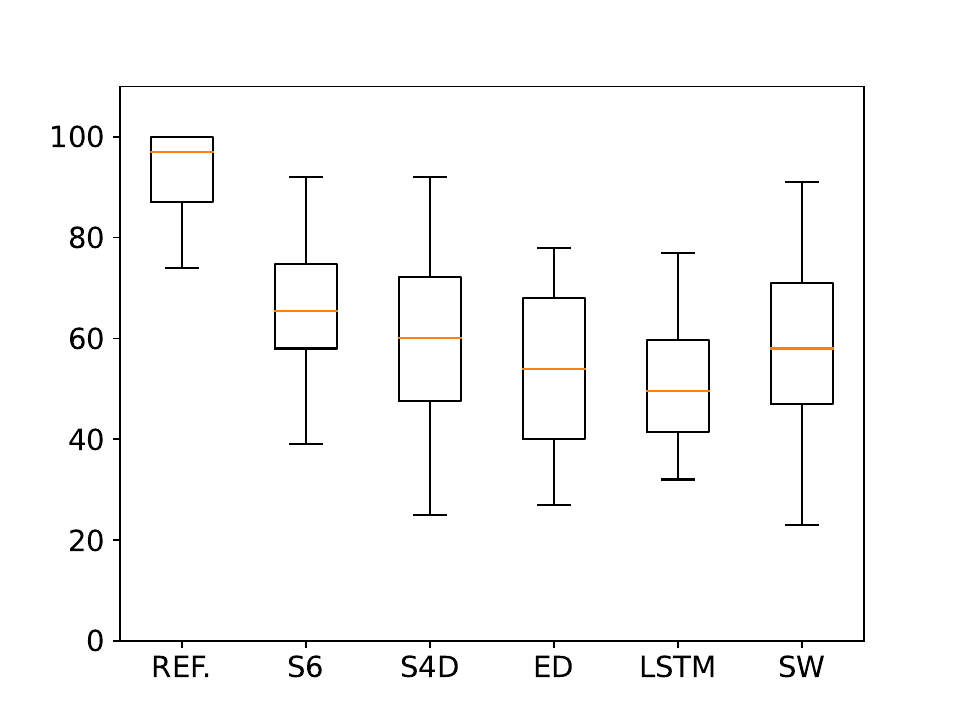}
\caption{Box plot displaying the MUSHRA rating results for the Teletronix LA-2A (top) and TubeTech CL 1B (bottom).}\label{fig:mushra}
\end{figure}

For the MUSHRA listening test, a total of $20$ individuals participated, with no participant meeting the criteria for result exclusion. The results are illustrated in Figure~\ref{fig:mushra}, where the box plots display the median, lower and upper quartiles, maximum and minimum values, and outliers of the responses. In the plots, the ratings for anchors have not been included to facilitate visual comparison of the models being evaluated. Results indicate that S6 models are slightly more perceptual accurate than others, offering the closest match to the reference. This is demonstrated by the median level and the narrower interquartile range, suggesting consistent performance across listening tests. The S4D models appear to be the second best, while the LSTM models performed particularly poorly when modeling the TubeTech CL 1B, likely due to noticeable audible artifacts. 

The ratings for compressed signals from the Universal Audio LA-2A and Softube CL 1B software emulations (SW), which are slightly lower than those of most models, require additional considerations. On one hand, the compression characteristics of these commercial products—details of which remain undisclosed—might differ from those produced by their actual analog counterparts, yet they are still considered sufficiently good for sound and music production. During the MUSHRA tests, participants were asked to rate the similarity to a reference signal. The complete absence of background noise in digital plugins, which is typically present in small amounts in recordings from analog devices, may have also contributed to their lower similarity ratings. Further studies are needed to confirm the validity and determine the respective weight of these hypotheses.

For further qualitative evaluation, we have computed a collection of plots to compare the response of the Teletronix LA-2A and TubeTech CL 1B, used as targets for the trainings, with that of the proposed S6 model and alternative models with recurrent blocks based on S4D, ED, and LSTM. These plots were generated for a selection of compression parameter settings representing cases commonly used in real application scenarios. Due to space limitations, this paper only includes those related to the following settings: Teletronix LA-2A in limit mode with peak reduction set to $20$ and compression mode with peak reduction set to $40$; TubeTech CL 1B with a threshold $-20$~dBu and ratio $2:1$, and with threshold $-10$~dBu and ratio $6:1$, both with attack and release set at $\sim 250$ ms and $\sim 5$ s. 

\begin{figure}[h]%
\centering
\includegraphics[width=0.22\textwidth]{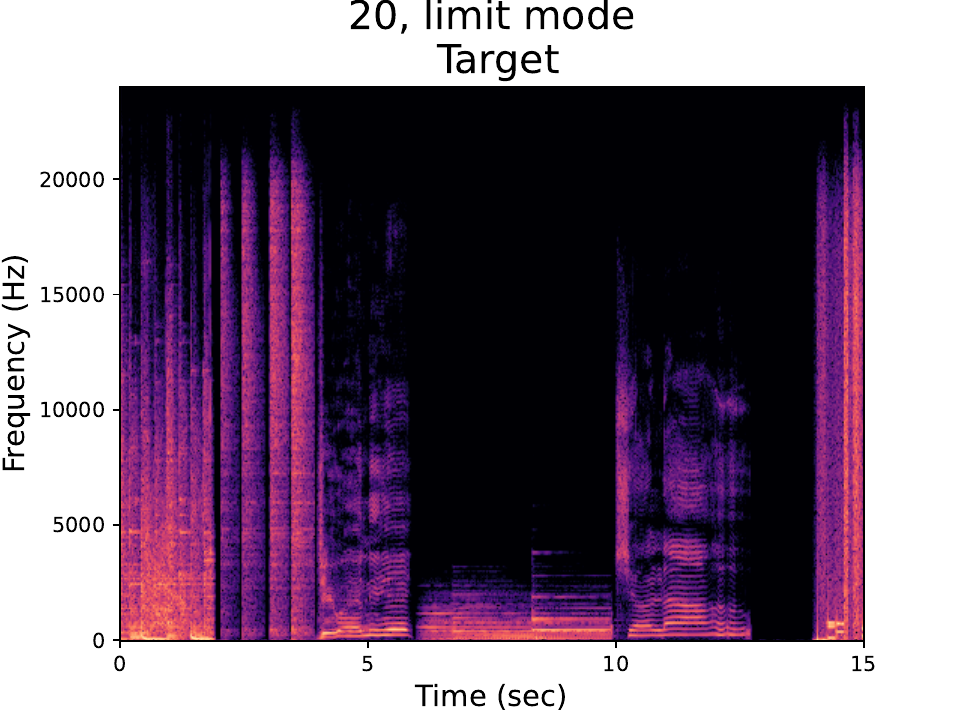}
\includegraphics[width=0.22\textwidth]{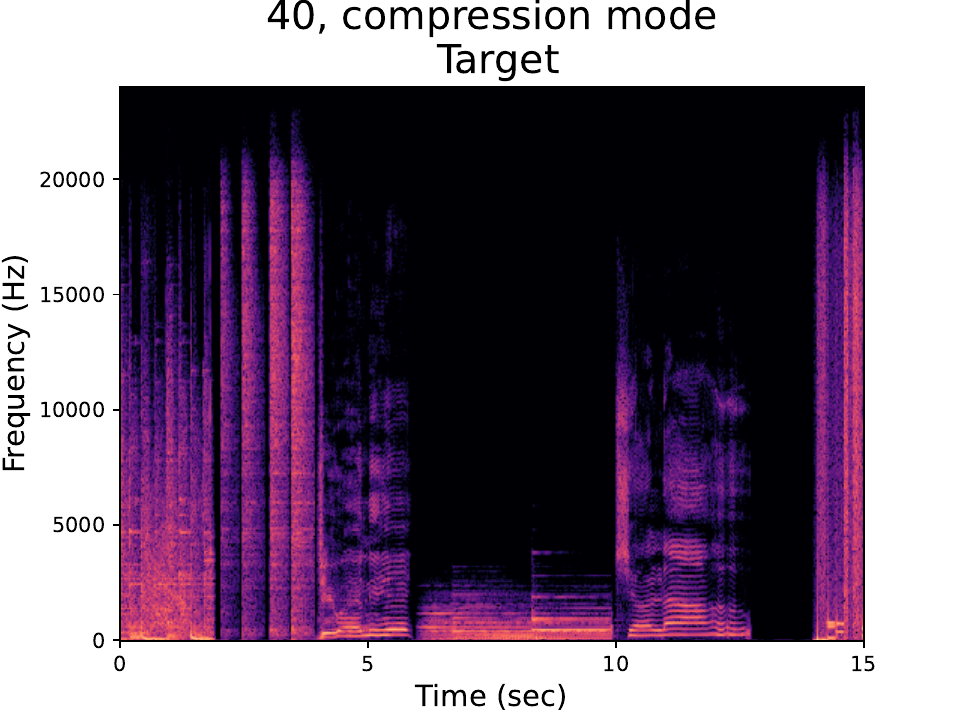}\\
\includegraphics[width=0.22\textwidth]{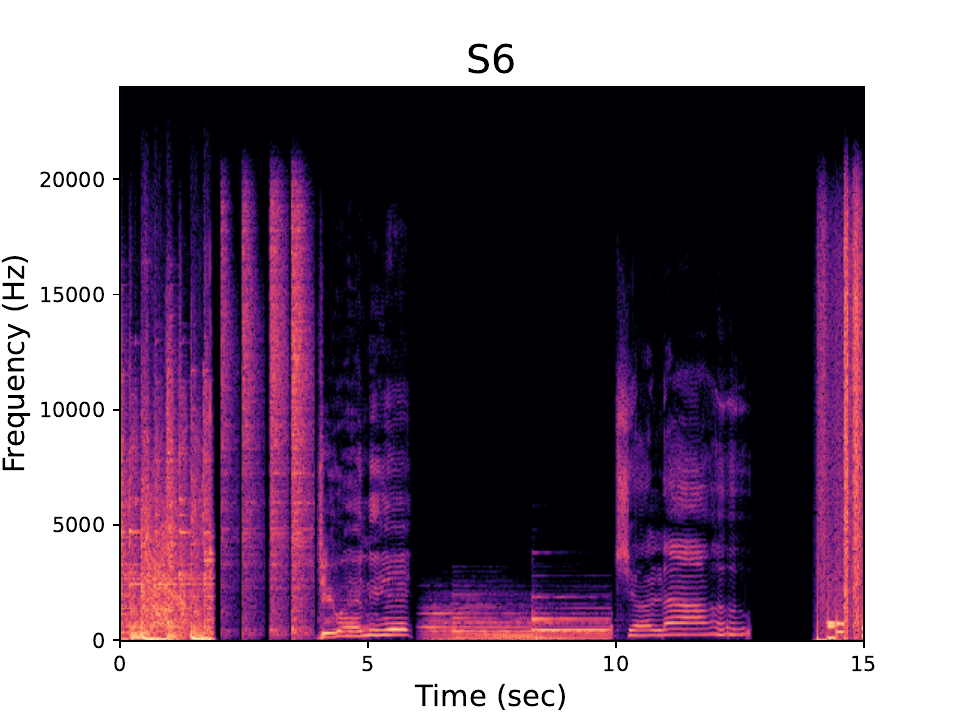}
\includegraphics[width=0.22\textwidth]{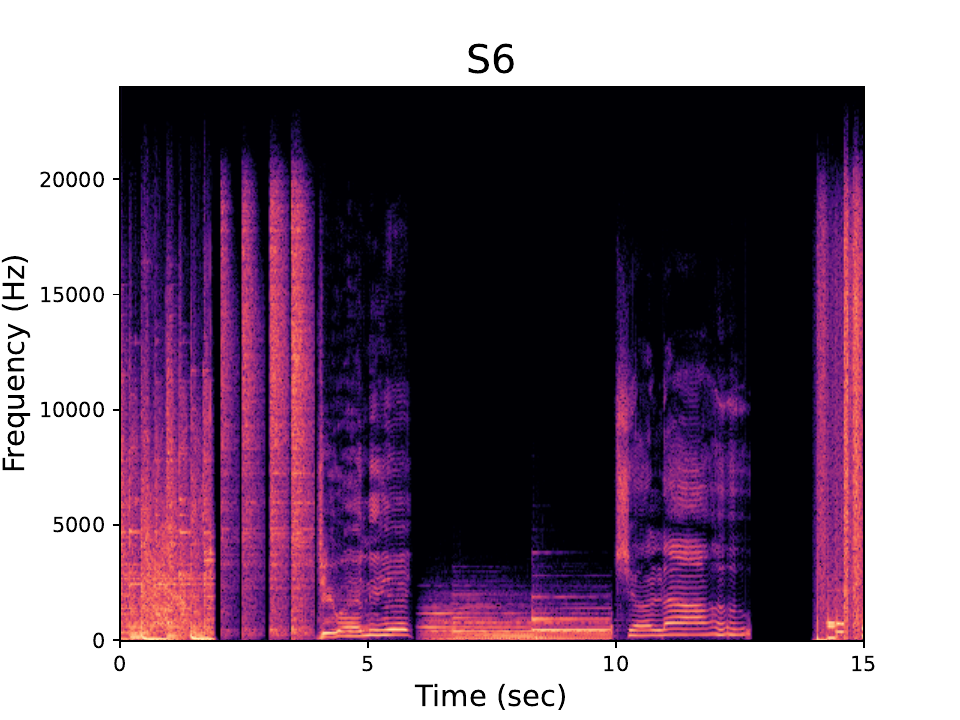}\\
\includegraphics[width=0.22\textwidth]{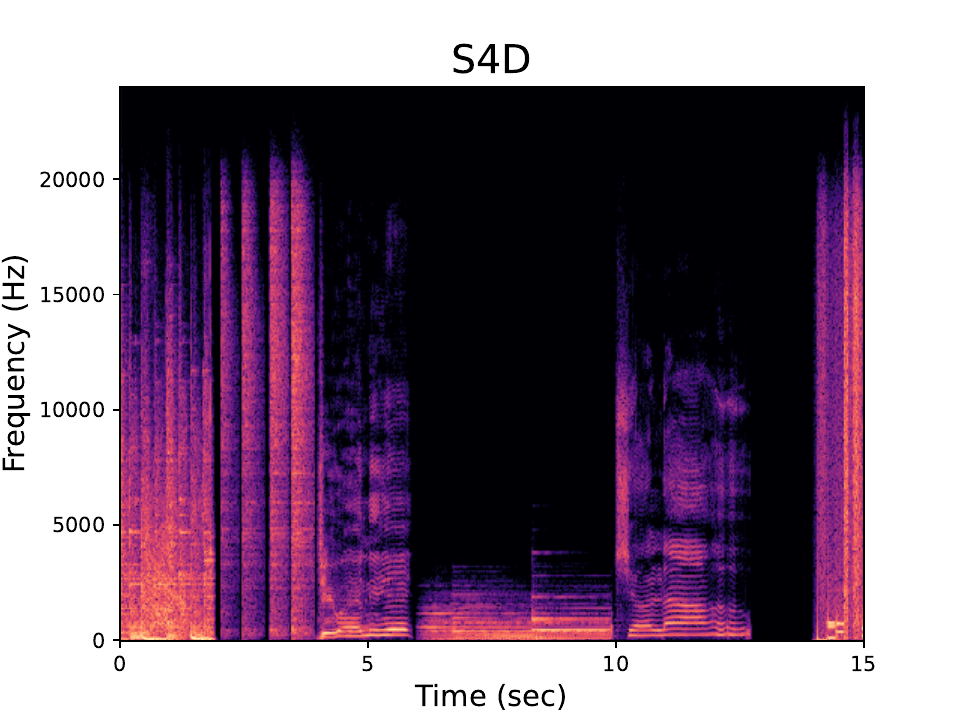}
\includegraphics[width=0.22\textwidth]{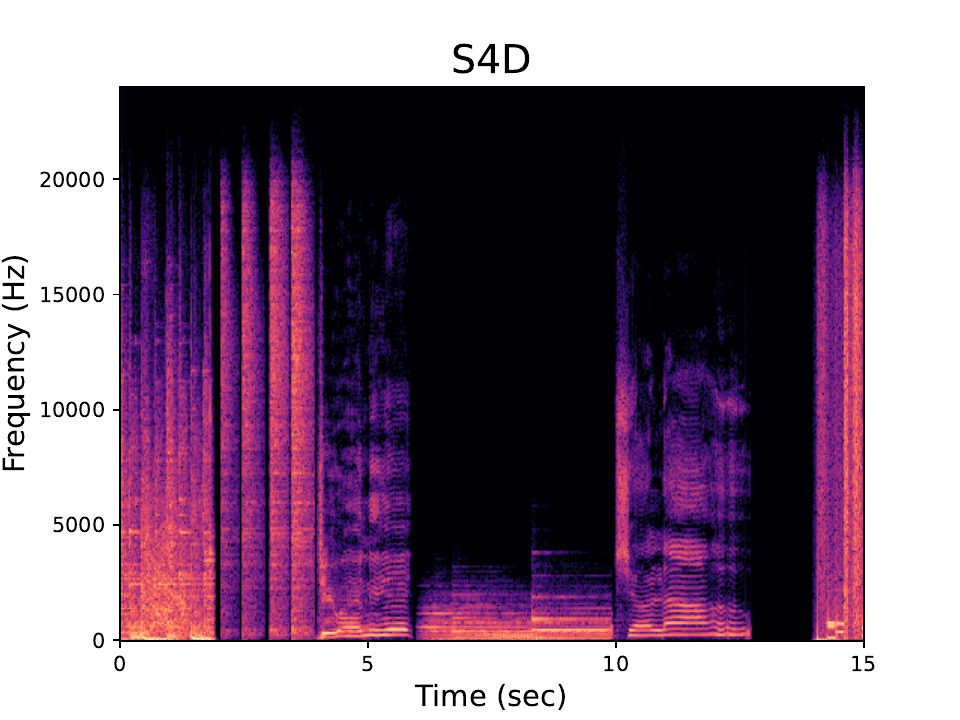}\\
\includegraphics[width=0.22\textwidth]{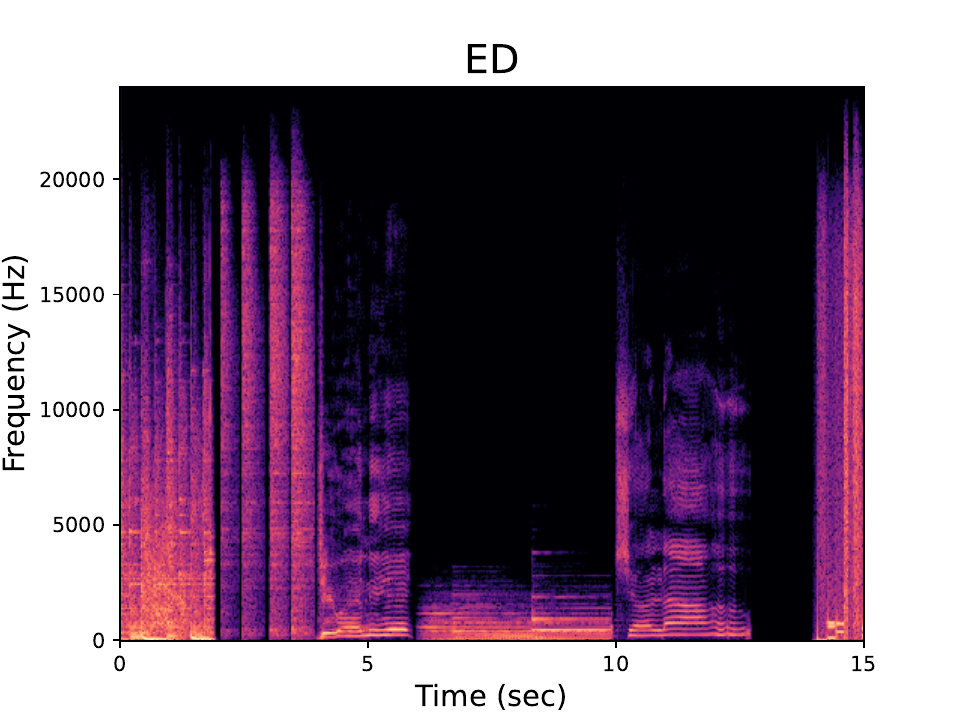}
\includegraphics[width=0.22\textwidth]{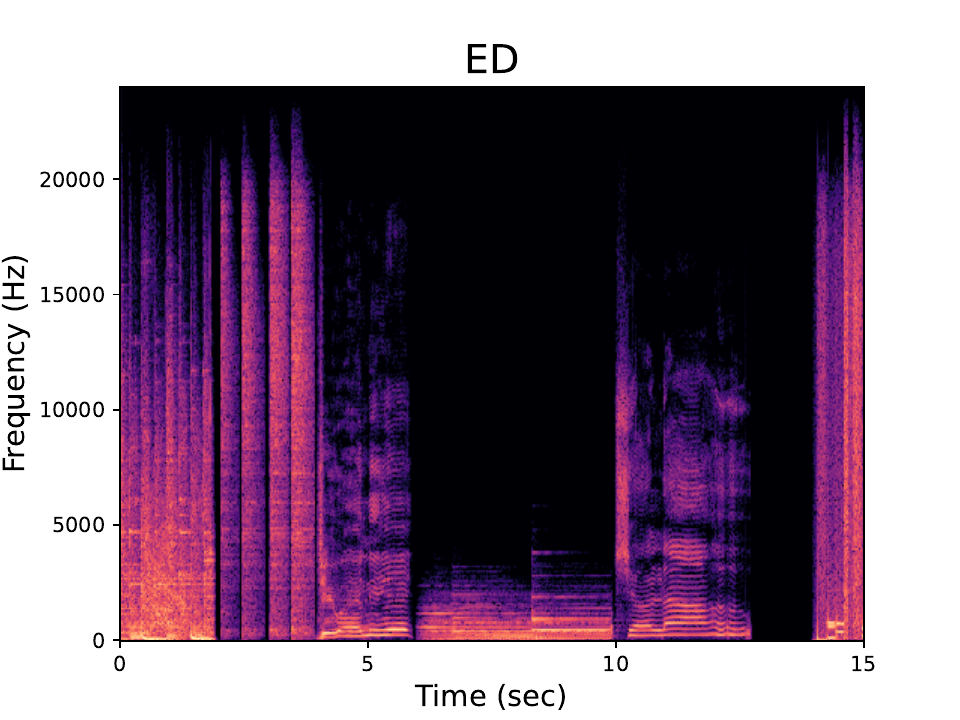}\\
\includegraphics[width=0.22\textwidth]{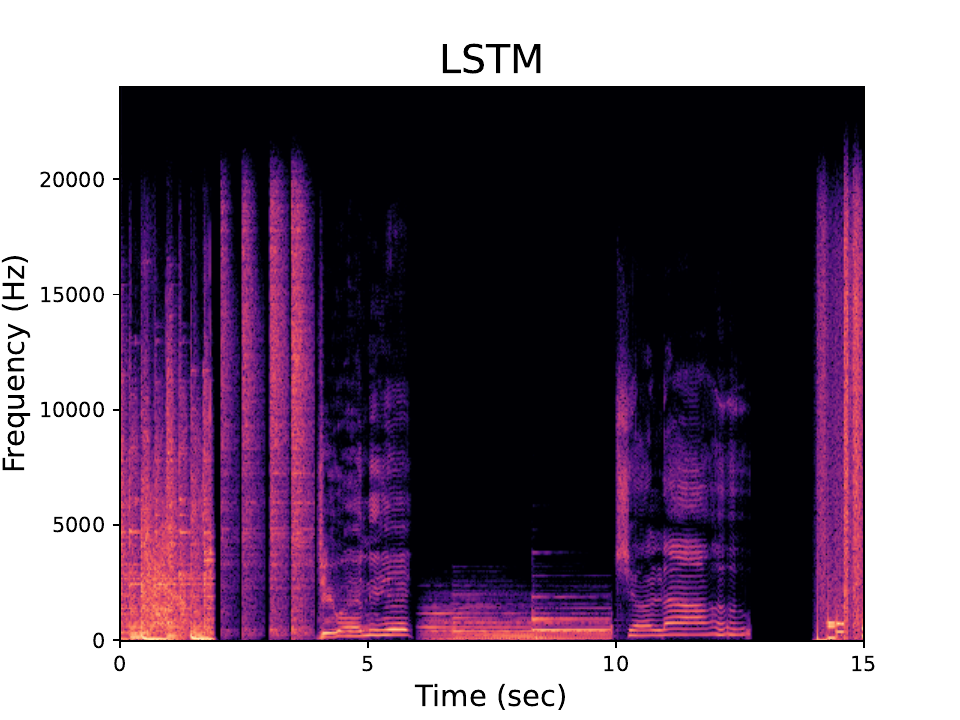}
\includegraphics[width=0.22\textwidth]{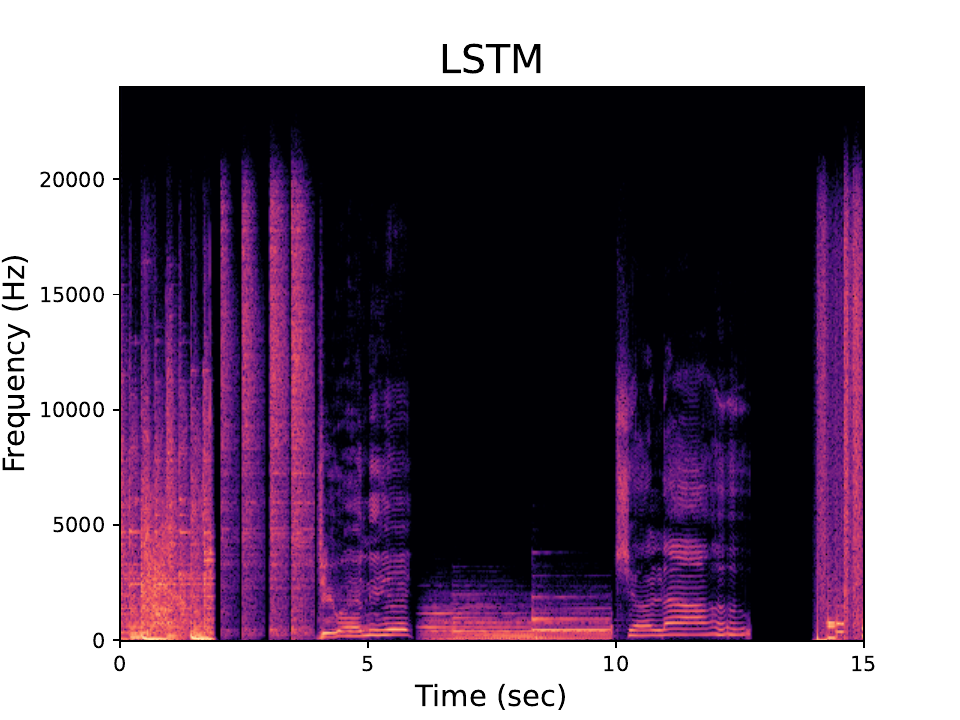}\\
\caption{Spectrograms of the output of the Teletronix LA-2A and of the output of models, trained using the proposed S6 and alternative architectures. Each column presents spectrograms for different compression settings labeled above the columns. The model that produced each spectrogram is specified at the top of the respective spectrogram. `Target' refers to the analog hardware device.}\label{fig:la_stft}
\end{figure}

\begin{figure}[h]%
\centering
\includegraphics[width=0.22\textwidth]{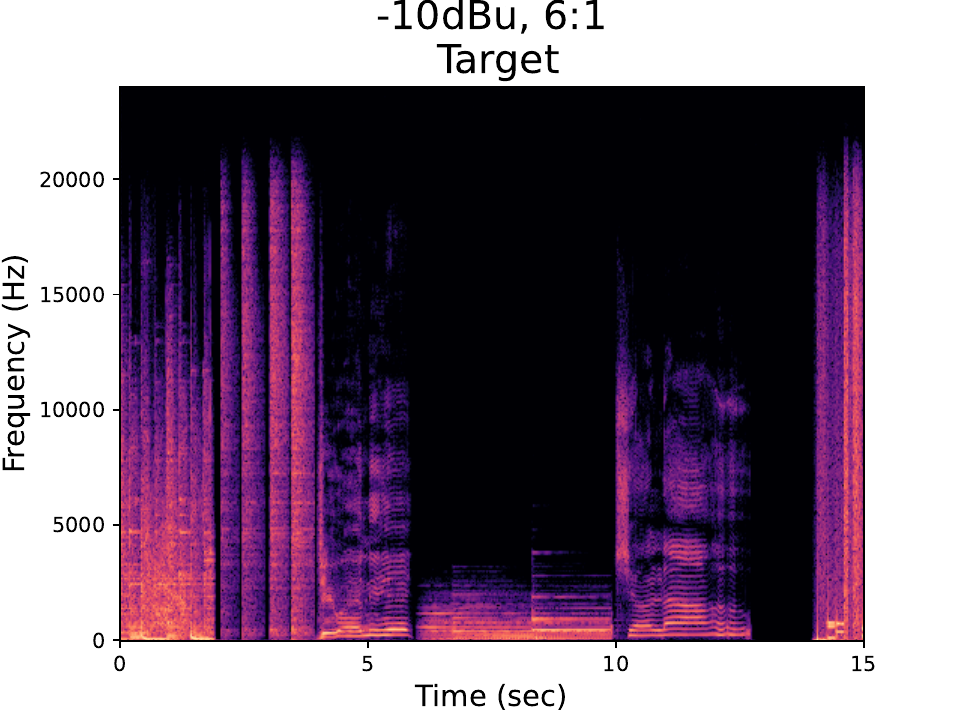}
\includegraphics[width=0.22\textwidth]{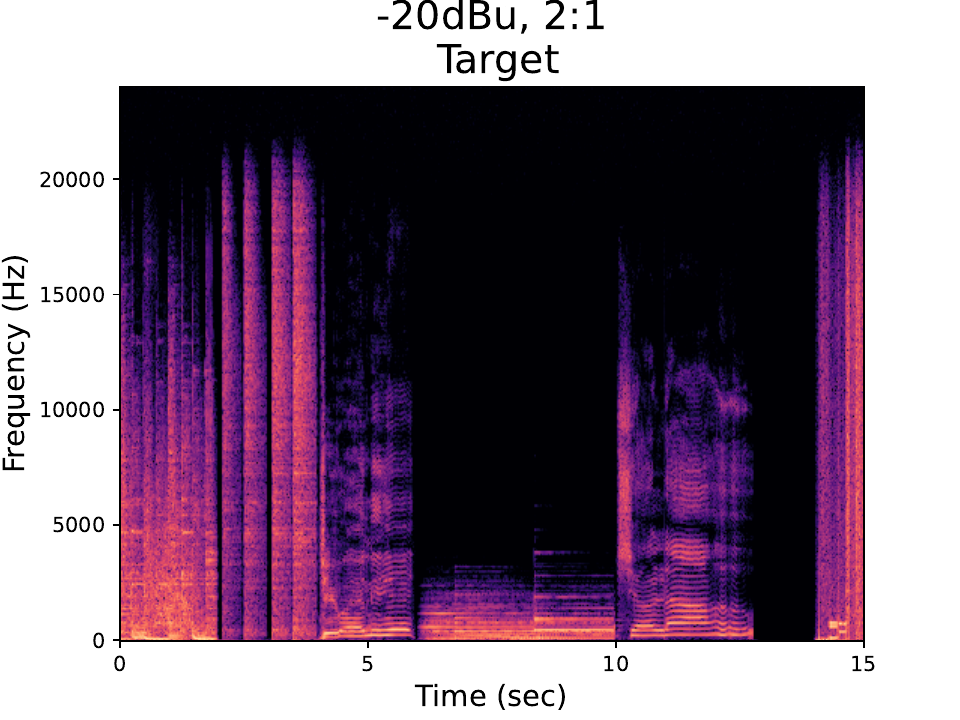}\\
\includegraphics[width=0.22\textwidth]{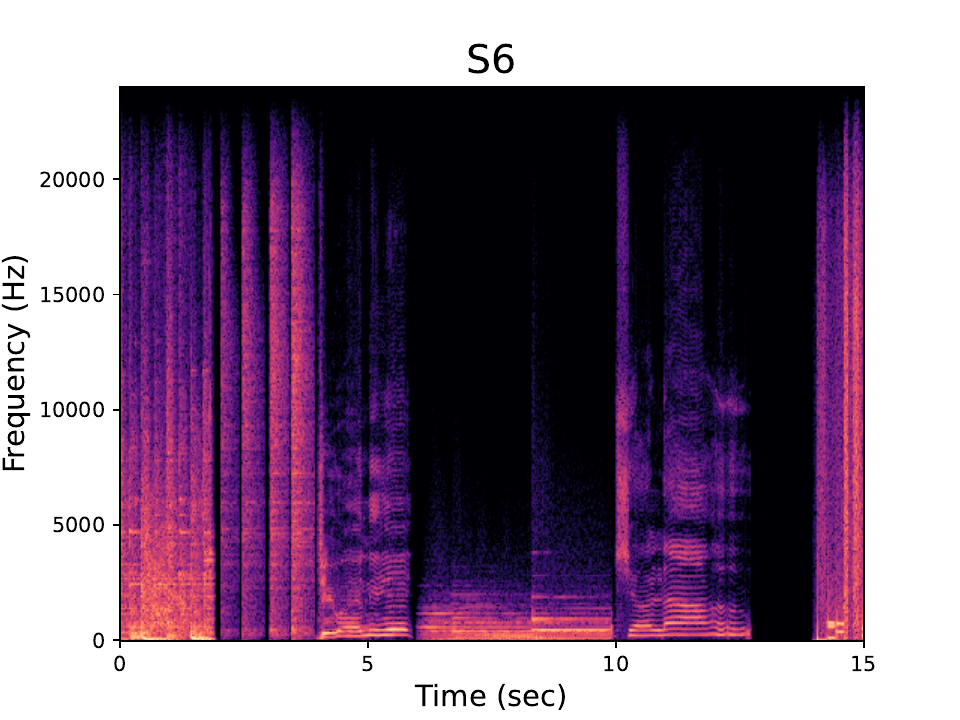}
\includegraphics[width=0.22\textwidth]{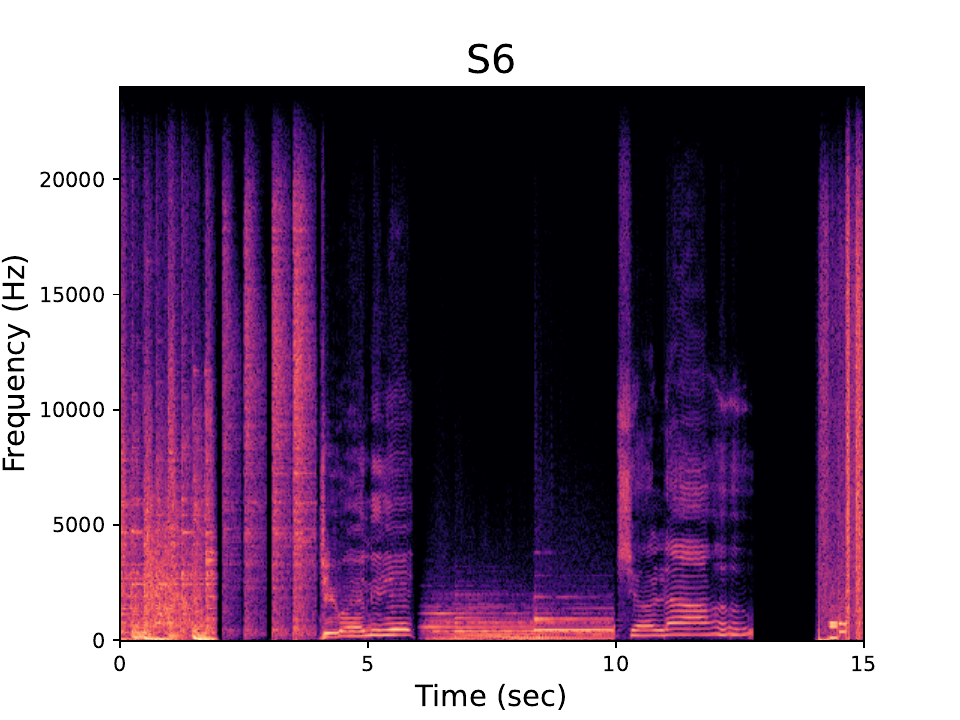}
\\
\includegraphics[width=0.22\textwidth]{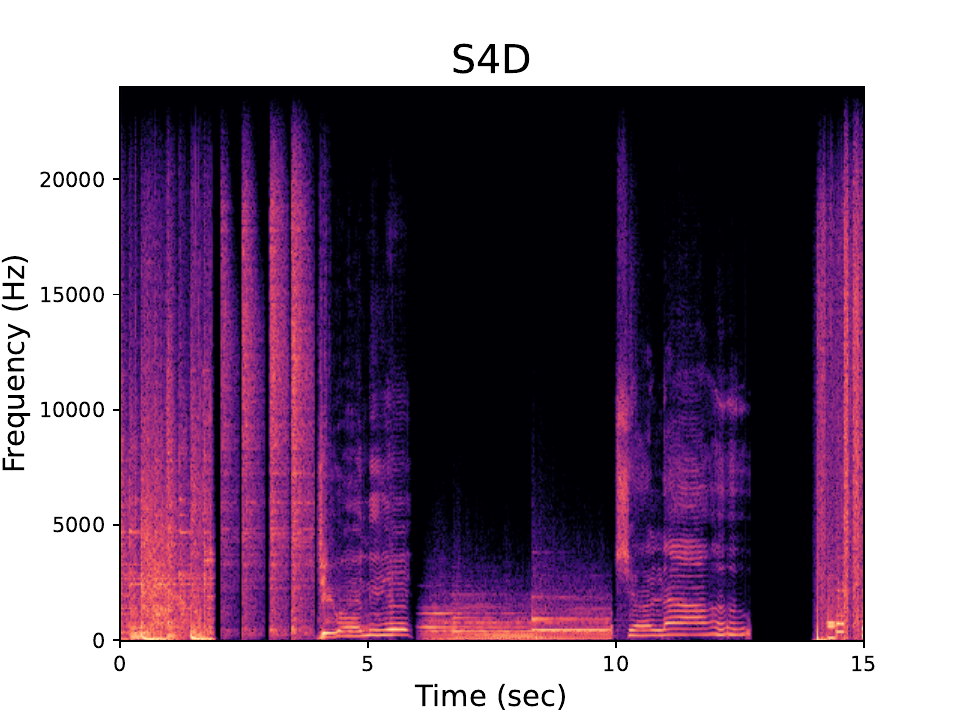}
\includegraphics[width=0.22\textwidth]{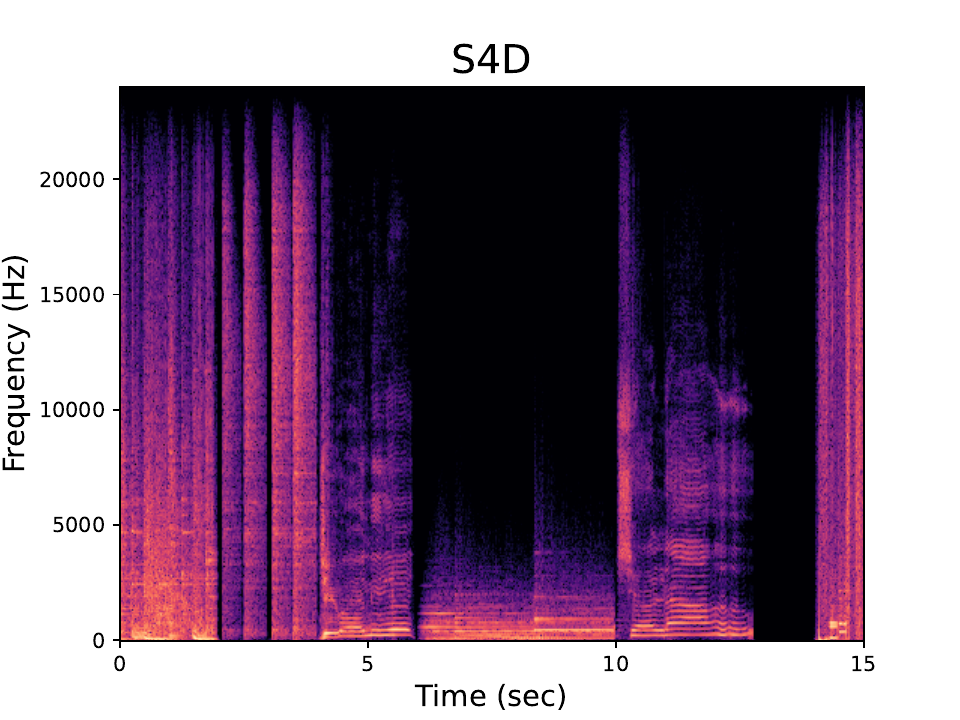}
\\
\includegraphics[width=0.22\textwidth]{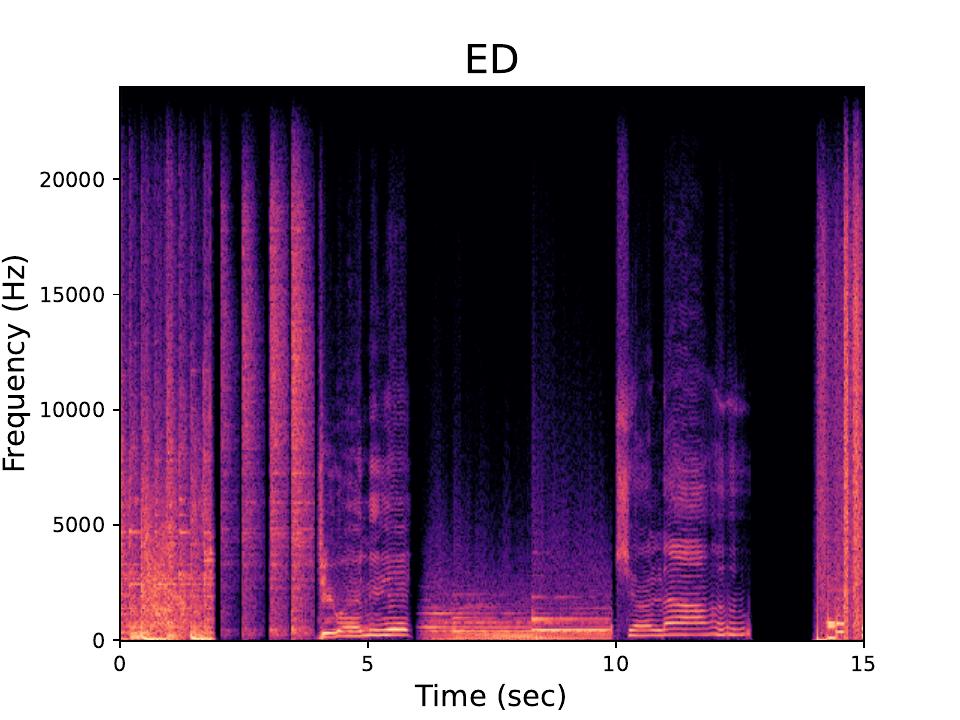}
\includegraphics[width=0.22\textwidth]{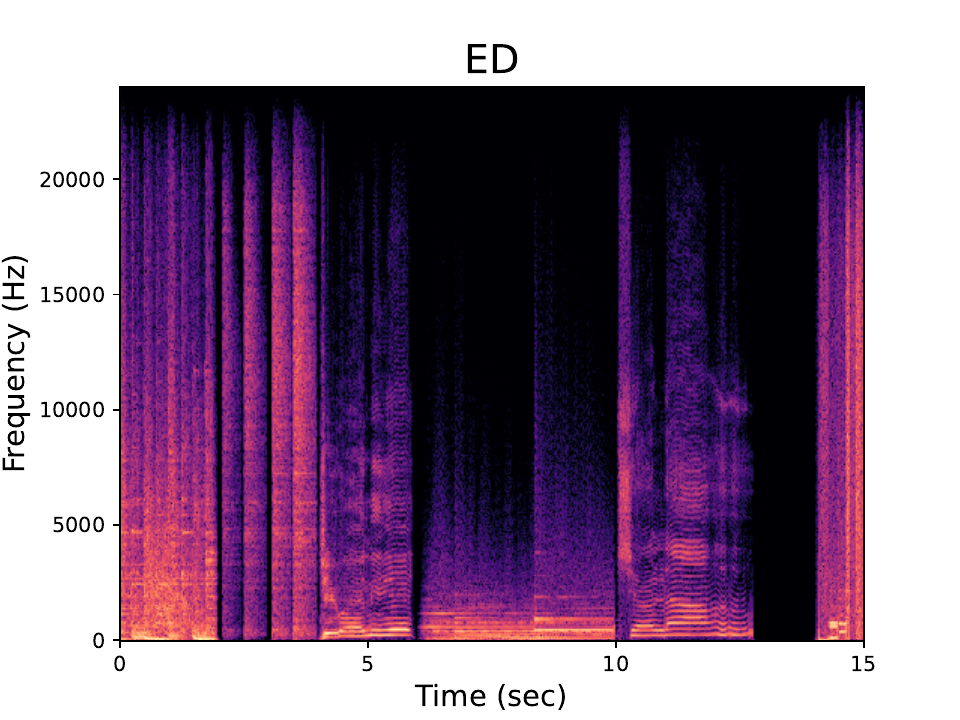}
\\
\includegraphics[width=0.22\textwidth]{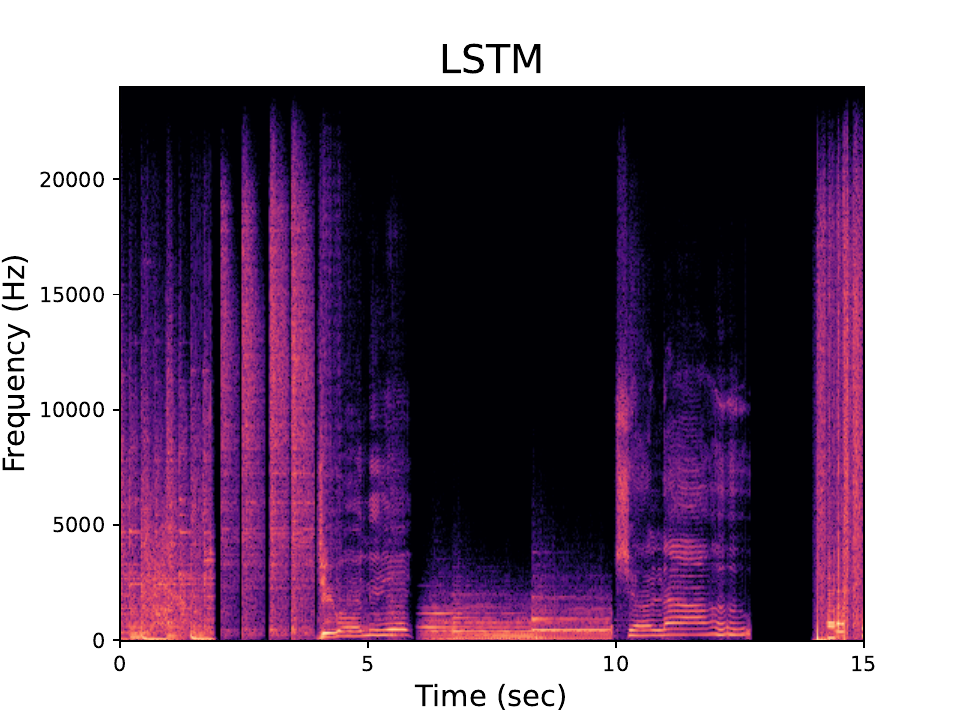}
\includegraphics[width=0.22\textwidth]{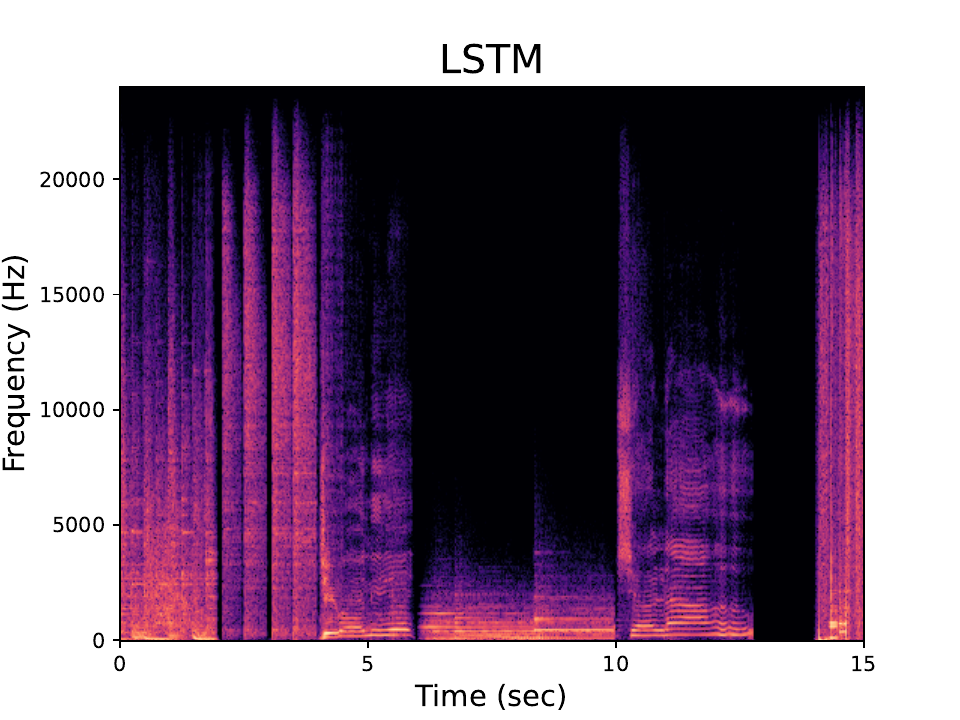}\\
\caption{Spectrograms of the output of the TubeTech CL 1B, and of the output of models trained using the proposed S6 and alternative architectures. Each column presents spectrograms for different compression settings labeled above the columns. The model that produced each spectrogram is specified at the top of the respective spectrogram. `Target' refers to the analog hardware device.}\label{fig:cl_stft}
\end{figure}

\begin{figure}[h]%
\centering
\includegraphics[width=0.45\textwidth]{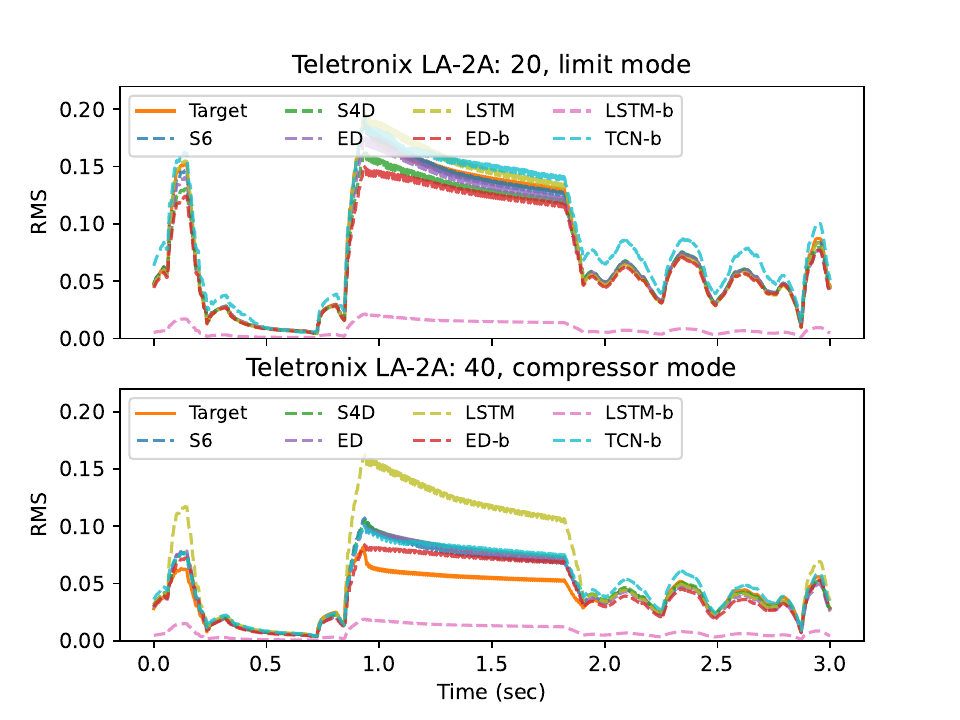}
\includegraphics[width=0.45\textwidth]{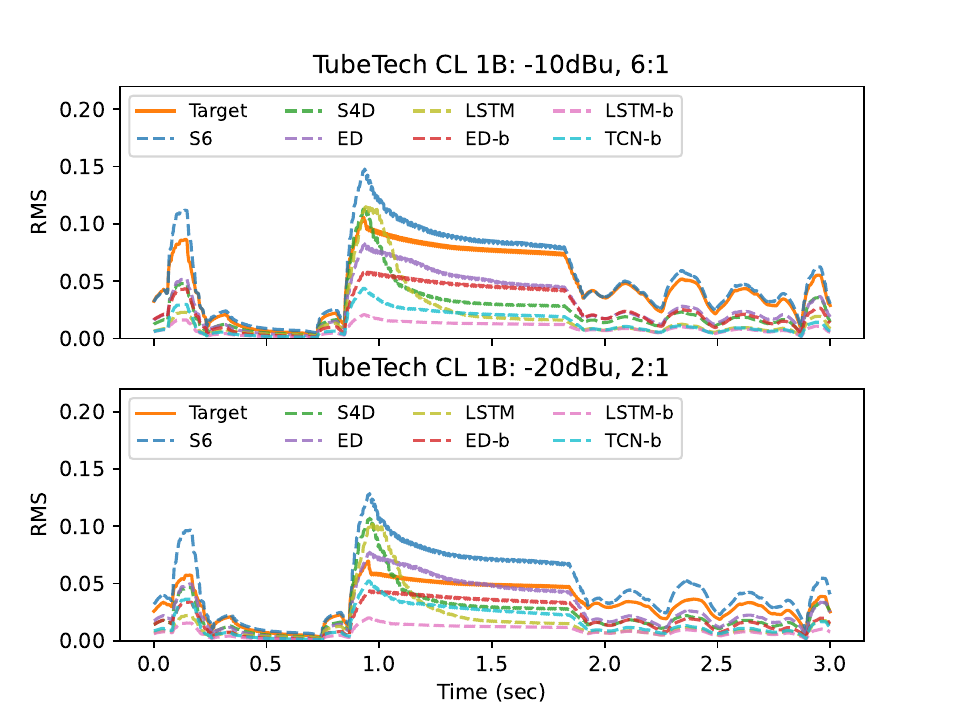}
\caption{RMS levels measured at the output of the Teletronix LA-2A (top) and TubeTech CL 1B (bottom) compared with those at the output of models trained using the proposed S6 and alternative architectures. The applied compression settings are indicated above the plots. The legend associates the different color curves with their corresponding models, with `Target' denoting the analog hardware device.}\label{fig:rmse}
\end{figure}

\begin{figure}[h]%
\centering
\includegraphics[width=0.45\textwidth]{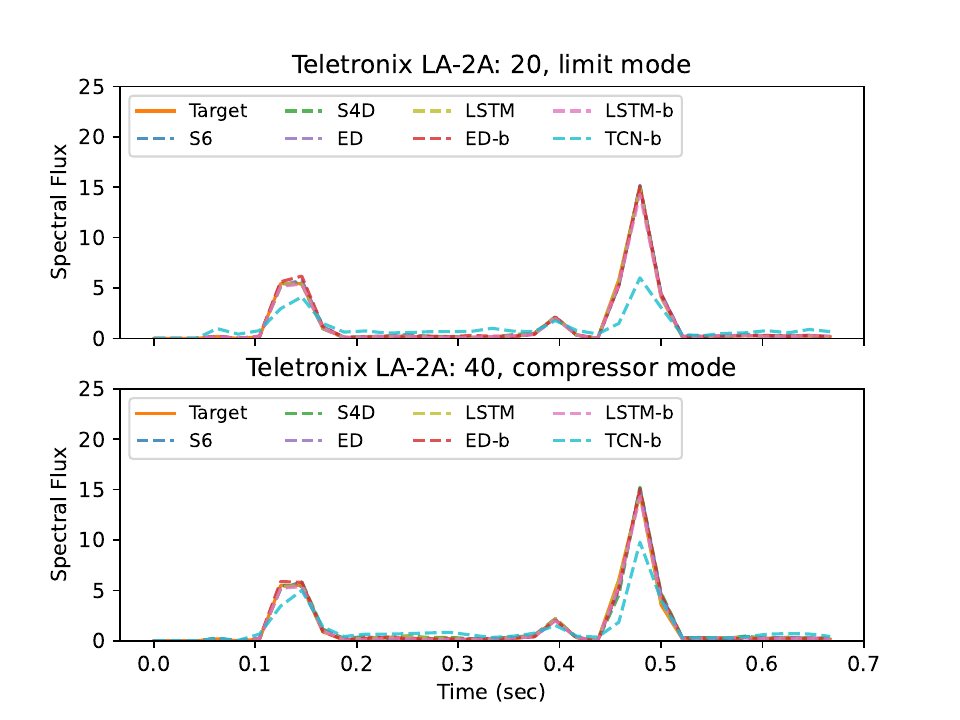}
\includegraphics[width=0.45\textwidth]{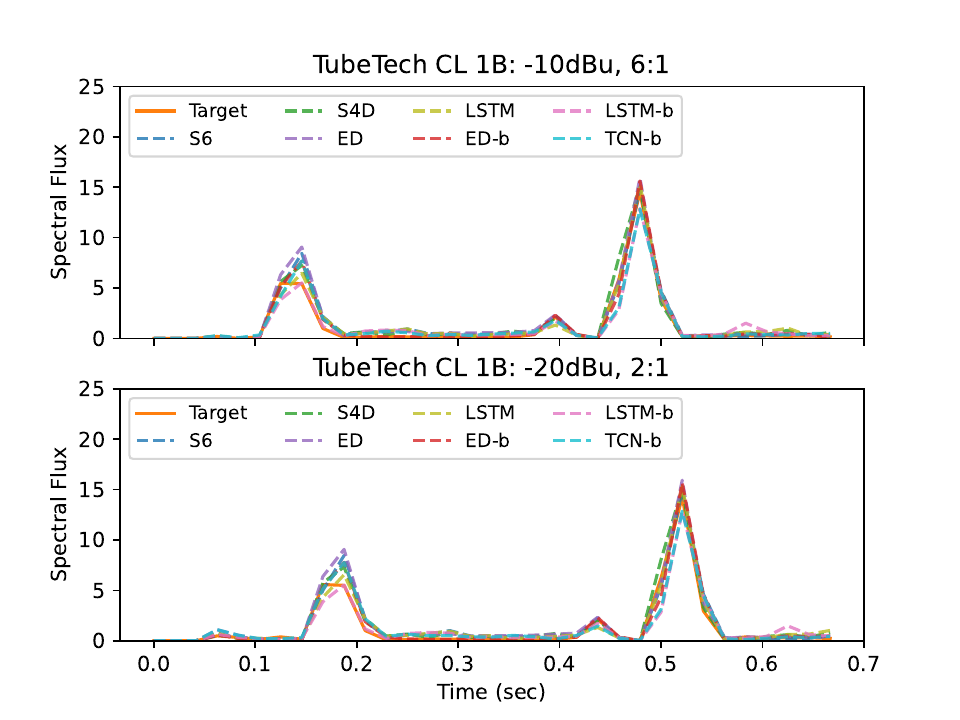}
\caption{Spectral flux measured at the output of the Teletronix LA-2A (top) and TubeTech CL 1B (bottom) compared with those at the output of models trained using the proposed S6 and alternative architectures. The applied compression settings are indicated above the plots. The legend associates the different color curves with their corresponding models, with `Target' denoting the analog hardware device.}\label{fig:sf}
\end{figure}

The plots we computed include the spectrograms of the output signals in Figure~\ref{fig:la_stft} and Figure~\ref{fig:cl_stft}, the RMS of the output providing an indication of the signal level over time in Figure~\ref{fig:rmse}, and the spectral flux of the output highlighting the response during sharp attacks in Figure~\ref{fig:sf}. Spectrograms are calculated using windows of $1024$ samples with $25\%$ overlap. The RMS and spectral flux are computed using windows of $4096$ samples with $25\%$ overlap.  Additional plots for more compression cases and the associated target and model-generated audio signals are available online~\footnote{\url{https://github.com/RiccardoVib/Optical-DRC-with-Selective-SSMs}}.

When looking at the spectrograms in Figure~\ref{fig:la_stft}, the difference appears less pronounced among the models showing a close matching, especially for S6 and S4D models. When considering the TubeTech CL 1B in Figure~\ref{fig:cl_stft}, the difference is more significant than the LA-2A case, showing more energy at a higher frequency than expected. This aspect is consistent with the higher M-STFTE in Table~\ref{tab:summaryanalog}. Here, the S6 and ED models present the best matches. 

In replicating the appropriate gain reduction required to match the target compressed signal, the LSTM also shows the poorest performance, as evident in the RMSE plot examples of Figure~\ref{fig:rmse}, which refers to a $3$-second snippet from the test set. The snippet features two plucked electric bass notes in the first two seconds, followed by vocals in the final second. The first bass note is muted, while the second, occurring in the middle of the snippet, has a longer sustain. These examples have been selected as they clearly illustrate the differing behaviors of various models compared to the target, especially during the sustained bass note, which presents the greatest challenge, representing a region of the test set with errors notably higher than the average. The concluding vocal segment displays a series of rapid, albeit less pronounced, level changes that several models fail to compress accurately. 

It appears that most models follow the amplitude envelope of the output signal, yet they vary in their degree of accuracy regarding the applied gain reduction. Additionally, the plot employs a linear amplitude scale; therefore, the mismatches' perceptibility does not directly correlate with their visual representation in the plots. The LSTM model does not compress the signal promptly in the LA-2A scenario, while the other model better emulates the timing behavior of the compression process, particularly evident when observing peak reduction at $40$. Meanwhile, the S6 model significantly outperforms other models with the mildest compression applied. Conversely, when looking at the CL 1B case, although the S6 model has better matching, all the models overcompress the signal, especially the models using LSTM and S4D in the recurrent block of the proposed architecture. The ED model does not properly reduce the dynamic but similarly presents the compression as constant.

The Spectral Flux graph in Figure~\ref{fig:sf}, computed for a $0.5$-second snippet from the test set two percussive sounds, reveals only minor differences among the models, indicating a consistent good match during the onsets. In the LA-2A case, TCN shows poor accuracy, while the other model shows an overall better emulation of the compression process during onsets. 

\subsection{Unseen settings and sampling density}

The results of the experiments to evaluate the ability of models to predict the compressor output for unseen parameter combinations confirm that the proposed S6 model performs better than alternative and baseline architectures. These experiments use the Universal Audio LA-2A, Softube CL 1B `dynamic,' and Softube CL 1B `time' datasets, assess the impact of the dataset size, with respect to the control parameter sampling density on the learning accuracy, as described in Section~\ref{sec:metrics}. The performance of the models is always evaluated on the same test set; results for each dataset are summarized in Table~\ref{tab:lacond}, Table~\ref{tab:din}, and Table~\ref{tab:time}, which detail the metrics across datasets of varying sizes due to differences in parameter sampling density.

The results for the LA-2A dataset confirm that the S6 model has better performance. When the dataset size increases, the S6 model accuracy increases as well, together with the gap between the other proposed models. Only the ED model presents slightly and negligible better errors for the smaller size. Noticeably, all other models resulted in higher errors compared to the smaller case. Indeed, having more examples of parameter combinations does not lead to lower errors for all the models. In particular, ED, S4D, and LSTM behaved better when trained with the fewest number of examples. Similar behavior is encountered with the \texttt{Softube CL 1B dynamic} dataset. 

The S6 model is not always ranked first, but the metrics are very close to the best ones in all the scenarios, and it achieves the best absolute accuracy with the medium-size dataset. ED again shows better predictions for unseen combinations in the smaller dataset. Still increasing, dataset S6 becomes more accurate than other models and than the previous case, which is the opposite of the other models. Also, in this case, the best accuracy is achieved with the medium-size dataset.

Finally, the case of unseen attack and release times show the opposite trend. This time, the errors are generally larger for all models when increasing the dataset size, as well as for the S6 model, which results in the best performing one also in this case. The bigger errors with this dataset confirm a difficult case when including time variant dependencies. 

In general, S6 is more consistent and stable when changing the dataset size. When considering parameters that influence the dynamics, S6's performance improves with an increased number of input-output examples; in contrast, a decline is observed when dealing with data characterized by variable temporal behavior. As mentioned, we do not know the scale of the parameter ranges. We sampled the parameters with equal-spaced steps, considering a linear range. In the case of logarithmic scales, we have more examples covering the part close to one extreme and a few examples for the other one, leading to over-fitting. Results also indicate that LSTM models, especially ED, are better when interpolating unseen parameters when a few examples are available. This suggests they better encode the input information when having a relatively small amount of examples. Conversely, S6 and S4D are better when more examples are seen during training. 

Lastly, bigger datasets do not always lead to lower error; the smallest ones sometimes present more accurate results. This means that too many examples can confuse the network, while fewer points in the space help better approximate the different conditioning cases overall.
\begin{table*}[t]
\centering
\tabcolsep8.1pt
\caption{Loss and performance metrics for models trained on \texttt{Universal Audio LA-2A} datasets with different parameter sampling densities. The `Set' columns indicate the training sets used, corresponding to the suffix of the training set names detailed in Table~\ref{tab:summarytraintest}. Models with the lowest loss and the minimum value for each metric are highlighted in bold.}
\label{tab:lacond}
{%
\begin{tabular}{@{}lllllllllc@{}}\toprule
Set & Model & MSE & MAE & ESR & RMSE & SFE & M-STFTE & PSM\\
\texttt{0.10} & S6 & $ 6.10 \cdot 10^{-4} $ & $\mathbf{1.19 \cdot 10^{-2}}$ & $ 1.17 \cdot 10^{-1}$ & $ 1.10\cdot 10^{-2}$ & $\mathbf{1.02\cdot 10^{-2}}$ & $1.86$ & $0.95$\\
& S4D & $8.04\cdot 10^{-4} $ & $ 1.34\cdot 10^{-2}$ & $ 1.54 \cdot 10^{-1}$ & $ 1.24 \cdot 10^{-2}$   & $ 1.33\cdot 10^{-2}$ & $ 2.21$ & $0.95$\\
& \textbf{ED} & $\mathbf{5.91 \cdot 10^{-4}}$ & $\mathbf{1.19 \cdot 10^{-2}}$ & $\mathbf{1.14\cdot  10^{-1}}$ & $\mathbf{1.09 \cdot 10^{-2}}$ & $ 1.37\cdot 10^{-2}$& $\mathbf{1.26}$  & $0.95$\\
& LSTM & $ 6.41 \cdot 10^{-4} $ & $ 1.20 \cdot 10^{-2}$ & $1.23 \cdot 10^{-1}$ & $1.11 \cdot 10^{-2}$  & $1.38\cdot 10^{-2}$ & $1.67$ & $0.95$\\
\hline
\texttt{0.05} & \textbf{S6} & $\mathbf{5.90 \cdot 10^{-4}} $ &$\mathbf{1.18\cdot 10^{-2}}$ & $\mathbf{1.13\cdot 10^{-1}}$ & $\mathbf{1.09 \cdot 10^{-2}}$ & $ 1.50\cdot 10^{-2}$ & $\mathbf{1.37}$ & $0.95$\\
& S4D & $1.09\cdot 10^{-3} $ & $1.54 \cdot 10^{-2}$ & $2.10 \cdot 10^{-1}$ & $ 1.44\cdot 10^{-2}$ & $\mathbf{1.10\cdot 10^{-2}}$ & $2.23$  & $0.95$\\
& ED & $7.42 \cdot 10^{-4}$ & $ 1.28\cdot 10^{-2}$ & $1.43 \cdot 10^{-1}$ & $ 1.21 \cdot 10^{-2}$ & $1.47 \cdot 10^{-2}$ & $1.86$ & $0.94$\\
& LSTM & $4.76  \cdot 10^{-3} $ & $ 3.67 \cdot 10^{-2}$ & $9.18 \cdot 10^{-1}$ & $3.66 \cdot 10^{-2}$ & $ 1.46\cdot 10^{-2}$ & $1.96$  & $0.90$\\
\hline
\texttt{0.01} & \textbf{S6} & $\mathbf{5.70 \cdot 10^{-4}} $ & $\mathbf{ 1.20 \cdot 10^{-2}}$ & $\mathbf{ 1.09\cdot 10^{-1}}$ & $\mathbf{1.12  \cdot 10^{-2}}$ & $ 1.03\cdot 10^{-2}$& $\mathbf{1.21}$  & $0.94$\\
& S4D & $4.68\cdot 10^{-3} $ & $ 3.60 \cdot 10^{-2}$ & $9.02 \cdot 10^{-1}$ & $ 3.58 \cdot 10^{-2}$ & $ 1.08\cdot 10^{-2}$& $ 2.98$  & $0.92$\\
& ED & $4.43 \cdot 10^{-3} $ & $ 3.42 \cdot 10^{-2}$ & $8.34 \cdot 10^{-1}$ & $3.39 \cdot 10^{-2}$ & $\mathbf{9.92\cdot 10^{-3}}$ & $2.97$ & $0.92$\\
& LSTM & $ 4.77 \cdot 10^{-3} $ & $3.66\cdot 10^{-2}$ & $9.19 \cdot 10^{-1}$ & $ 3.64\cdot 10^{-2}$  & $ 1.25\cdot 10^{-2}$& $2.98$  & $0.90$\\
\end{tabular}}
\end{table*}
%

\begin{table*}[t]
\centering
\tabcolsep8.1pt
\caption{Loss and performance metrics for models trained on \texttt{Softube CL 1B dynamic} datasets with different parameter sampling densities. The `Set' columns indicate the training sets used, corresponding to the suffix of the training set names detailed in Table~\ref{tab:summarytraintest}. Models with the lowest loss and the minimum value for each metric are highlighted in bold.}
\label{tab:din}
{%
\begin{tabular}{@{}lllllllllc@{}}\toprule
Set & Model & MSE & MAE & ESR & RMSE & SFE & M-STFTE & PSM\\
\texttt{0.20} & S6 & $ 1.05 \cdot 10^{-4} $ & $ 6.80 \cdot 10^{-3}$ & $ 2.12 \cdot 10^{-1}$ & $ 6.50 \cdot 10^{-3}$ & $\mathbf{1.02\cdot 10^{-2}}$& $ 1.86$ & $0.94$\\
& S4D & $1.60 \cdot 10^{-4} $ & $ 8.41 \cdot 10^{-3}$ & $ 3.25\cdot 10^{-1}$ & $ 8.07 \cdot 10^{-3}$ & $ 1.33\cdot 10^{-2}$& $2.21$ & $0.93$\\
& \textbf{ED} & $ \mathbf{6.99\cdot 10^{-5}} $ & $\mathbf{ 5.55 \cdot 10^{-3}}$ & $ \mathbf{1.41\cdot 10^{-1}}$ & $\mathbf{5.22 \cdot 10^{-3}}$  & $1.37 \cdot 10^{-2}$& $\mathbf{1.26}$ & $0.92$\\        
& LSTM & $ 9.29 \cdot 10^{-5} $ & $ 6.27 \cdot 10^{-3}$ & $1.87 \cdot 10^{-1}$ & $5.97 \cdot 10^{-3}$  & $ 1.38\cdot 10^{-2}$& $1.67$ & $0.91$\\
\hline
\texttt{0.10} & \textbf{S6} & $ \mathbf{5.41 \cdot 10^{-5}} $ & $ \mathbf{4.81\cdot 10^{-3}}$ & $\mathbf{1.09 \cdot 10^{-1}}$ & $ \mathbf{4.56\cdot 10^{-3}}$ & $1.30\cdot 10^{-2}$ & $\mathbf{1.14} $ & $0.92$\\
& S4D & $8.74 \cdot 10^{-5} $ & $6.27 \cdot 10^{-3}$ & $1.76\cdot 10^{-1}$ & $ 5.91 \cdot 10^{-3}$ & $\mathbf{9.69\cdot 10^{-3}}$ & $1.51$ & $0.96$\\
& ED & $ 7.60 \cdot 10^{-5} $ & $5.73 \cdot 10^{-3}$ & $ 1.53\cdot 10^{-1}$ & $5.47 \cdot 10^{-3}$  & $ 1.27\cdot 10^{-2}$ & $ 1.56$ & $0.94$\\
& LSTM & $ 1.75 \cdot 10^{-4} $ & $ 9.06  \cdot 10^{-3}$ & $ 3.53 \cdot 10^{-1}$ & $ 8.87 \cdot 10^{-3}$ & $1.62\cdot 10^{-2}$ & $2.36$ & $0.90$\\
\hline
\texttt{0.05} & \textbf{S6} & $ \mathbf{6.46 \cdot 10^{-5} }$ & $\mathbf{5.28  \cdot 10^{-3}}$ & $ \mathbf{1.30 \cdot 10^{-1}}$ & $\mathbf{5.03 \cdot 10^{-3}}$& $1.50 \cdot 10^{-2}$ & $\mathbf{1.37}$ & $0.92$\\
& S4D & $1.57 \cdot 10^{-4} $ & $ 8.46\cdot 10^{-3}$ & $3.17 \cdot 10^{-1}$ & $8.18 \cdot 10^{-3}$ & $\mathbf{1.10\cdot 10^{-2}}$ & $2.23$ & $0.94$\\
& ED & $ 1.02\cdot 10^{-4} $ & $6.77 \cdot 10^{-3}$ & $ 2.08 \cdot 10^{-1}$ & $6.58\cdot 10^{-3}$ & $ 1.47\cdot 10^{-2}$& $ 1.86$ & $0.92$ \\
& LSTM & $ 1.12 \cdot 10^{-4} $ & $ 7.16\cdot 10^{-3}$ & $2.27\cdot 10^{-1}$ & $6.87 \cdot 10^{-3}$ & $ 1.46\cdot 10^{-2}$& $ 1.96$ & $0.92$\\
\end{tabular}}
\end{table*}


\begin{table*}[t!]
\centering
\tabcolsep8.1pt
\caption{Loss and performance metrics for models trained on \texttt{Softube CL 1B time} datasets with different parameter sampling densities. The `Set' columns indicate the training sets used, corresponding to the suffix of the training set names detailed in Table~\ref{tab:summarytraintest}. Models with the lowest loss and the minimum value for each metric are highlighted in bold.}
\label{tab:time}
{%
\begin{tabular}{@{}lllllllllc@{}}\toprule
Set & Model & MSE & MAE & ESR & RMSE & SFE & M-STFTE & PSM\\
\texttt{0.20} & \textbf{S6} & $ \mathbf{3.32  \cdot 10^{-4}} $ & $ \mathbf{8.83 \cdot 10^{-3}}$ & $\mathbf{ 2.43 \cdot 10^{-1}}$ & $ 8.48  \cdot 10^{-3}$ & $\mathbf{7.18\cdot 10^{-3}}$& $2.00$ & $0.94$\\
& S4D & $3.49\cdot 10^{-4} $ & $ 8.85 \cdot 10^{-3}$ & $ 2.55 \cdot 10^{-1}$ & $ 8.20  \cdot 10^{-3}$  & $ 7.73\cdot 10^{-3}$& $\mathbf{1.93}$ & $0.96$\\
& ED & $ 3.52  \cdot 10^{-4} $ & $ 8.36  \cdot 10^{-3}$ & $ 2.57  \cdot 10^{-1}$ & $\mathbf{ 7.94 \cdot 10^{-3}}$& $ 8.63\cdot 10^{-3}$& $1.95$ & $0.95$\\
& LSTM & $ 6.83 \cdot 10^{-4} $ & $ 1.27 \cdot 10^{-2}$ & $5.31 \cdot 10^{-1}$ & $1.23 \cdot 10^{-2}$& $ 8.53\cdot 10^{-3}$& $3.31$ & $0.95$\\
\hline
\texttt{0.10} & \textbf{S6} & $\mathbf{ 4.16  \cdot 10^{-4}}$ & $\mathbf{ 9.27 \cdot 10^{-3}}$ & $\mathbf{2.65\cdot 10^{-1}}$ & $ 8.90  \cdot 10^{-3}$ & $9.38\cdot 10^{-3}$& $2.11$ & $0.94$\\
& S4D & $5.78 \cdot 10^{-4}$ & $1.07 \cdot 10^{-2}$ & $ 3.68 \cdot 10^{-1}$ & $ 1.01  \cdot 10^{-2}$& $\mathbf{ 8.73\cdot 10^{-3}}$& $2.39$ & $0.96$\\
& ED & $ 4.32 \cdot 10^{-4}$ & $ 9.15  \cdot 10^{-2}$ & $ 2.75  \cdot 10^{-1}$ & $ 8.78 \cdot 10^{-3}$& $ 1.04\cdot 10^{-2}$& $2.14$ & $0.95$\\
& LSTM & $ 4.37 \cdot 10^{-4} $ & $ 9.14 \cdot 10^{-3}$ & $ 2.78\cdot 10^{-1}$ & $\mathbf{8.61 \cdot 10^{-3}}$& $ 9.42\cdot 10^{-3}$ & $\mathbf{2.01}$ & $ 0.95$ \\
\hline
\texttt{0.05} & S6 & $5.27 \cdot 10^{-4}$ & $1.10 \cdot 10^{-2}$ & $3.86  \cdot 10^{-1}$ & $1.07\cdot 10^{-2}$& $1.20 \cdot 10^{-2}$& $2.62$ & $0.92$\\
& \textbf{S4D} & $\mathbf{3.71 \cdot 10^{-4}}$ & $8.87\cdot 10^{-3}$ & $\mathbf{2.71 \cdot 10^{-1}}$ & $8.40  \cdot 10^{-3}$ & $7.98\cdot 10^{-3}$& $\mathbf{2.01}$ & $0.95$\\
& ED & $ 4.33 \cdot 10^{-4} $ & $\mathbf{8.78  \cdot 10^{-3}}$ & $3.17  \cdot 10^{-1}$ & $ \mathbf{8.31\cdot 10^{-3}}$& $ 9.22\cdot 10^{-3}$& $2.11$ &  $0.95$\\
& LSTM & $ 3.99  \cdot 10^{-4}$ & $ 8.97 \cdot 10^{-3}$ & $ 2.92\cdot 10^{-1}$ & $8.44 \cdot 10^{-3}$& $ \mathbf{7.71\cdot 10^{-3}}$& $2.07$ & $0.95$ \\
\end{tabular}}
\end{table*}

\subsection{Time behavior}

\begin{figure*}[h!]%
\centering
\includegraphics[width=1\textwidth]{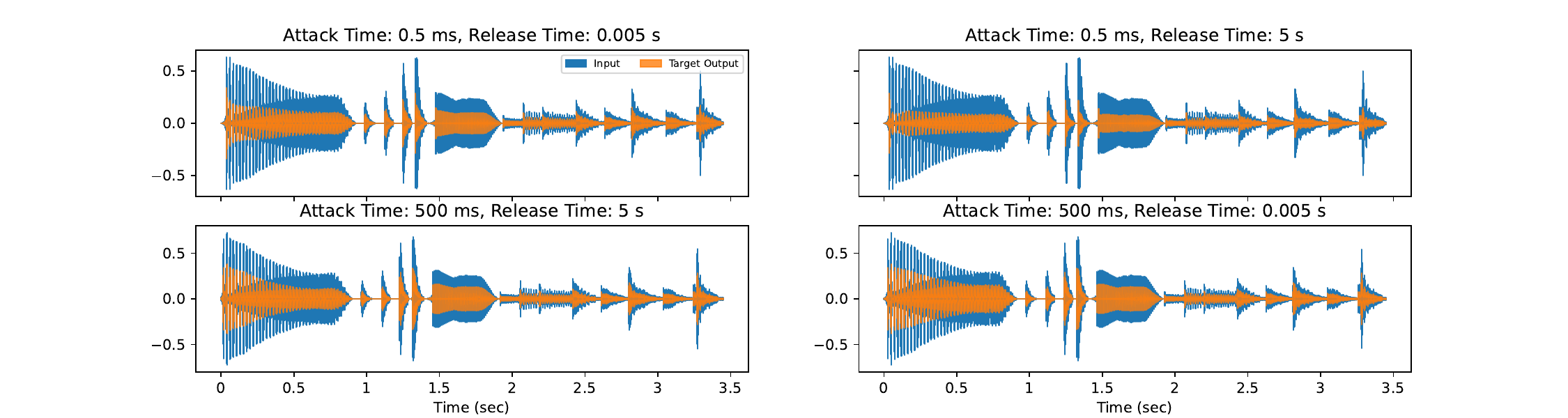}\\
\caption{Waveforms of input and output of the TubeTech CL 1B for the $3.60$-second signal, including bass and drum sounds, utilized as the test set in the experiments investigating the emulation of four different compression temporal profiles. The threshold and ratio are set to $-10$~dBu and $6:1$, respectively, while associated attack and release times are indicated above each plot.}\label{fig:TV}
\end{figure*}

Table~\ref{tab:TV} compares the loss and metrics for the models trained on four selected parameter combinations, featuring various settings of fast and/or slow attack and release times. The models are trained on individual compression settings, as well as on the combined set of four selected settings, as described in Section~\ref{sec:experiments}. In this context, given the short duration of the test signal and the significant differences in the local level of the compressed signal, the ESR metric is the most informative. It provides immediate information because it evaluates the ratio of error to signal level rather than the absolute value of the error. The impact of the different timing settings is visible in the waveform displayed in Figure~\ref{fig:TV}, which shows the selected test signal at both the input and output of the TubeTech CL 1B compressor for the four different compression settings.

Results indicate that the combination of fast attack and slow release is the most challenging compression scenario to emulate, yielding the highest errors across all proposed models trained on individual compression settings. On the other hand, slow attack scenarios pose significant challenges for the baseline models. In addition, the performance degrades markedly when comparing models trained exclusively on settings including slow attack or slow release to models trained on the four combinations. As observed in previous experiments, these outcomes suggest that learning variable attack times across the wide range of CL 1B is especially difficult. 

While all models show varying degrees of increased and decreased errors depending on the conditioning, they generally learn the conditioning cases to different extents, and the learning process seems to average out accuracy among them. Notably, S6 and LSTM models conditioned to handle multiple compression timing behaviors do not always perform worse; instead, they distribute errors across the examples, performing poorer in some settings and better in others. In particular, they improve the accuracy when both parameters are set to slow times or both are set to fast ones but decrease for the other conditions. 

The ED and S4D variations of the proposed architecture exhibit greater susceptibility to degradation, resulting in higher errors across all settings. In comparison, the baseline models generally demonstrate inferior performance, presenting more difficulties for slow attack settings than for fast ones. The proposed architecture with the S6 block consistently delivers the best performance, whereas the TCN-b model performs notably poorly, indicating a potential difficulty in learning varied timing behaviors and a tendency to favor for settings with rapid attack and release times. The low error rates might also be influenced by the limited dataset used for training, given the model's reliance on convolutional layers.
\begin{table*}[t!]
\centering
\tabcolsep8.1pt
\caption{Average loss and performance metrics obtained over five training iterations using a subset of the \texttt{TubeTech CL 1B time} dataset. The compression parameters of the selected recording are threshold and ratio fixed to $-10$~dBu and $6:1$, attack times of $0.5$~ms ($a_f$) and~$500$ ms ($a_s$), and release times of $0.005$~s ($r_f$) and $5$~s ($r_s$). The left half of the table corresponds to models trained on a single compression setting, whereas the right half pertains to models trained across all four timing variations of the compression settings. The top row for each model in the right section summarizes the loss and metrics across all parameter combinations. The symbols $\uparrow$ and $\downarrow$ on the right side signify an increase (worse performance) or decrease (improved performance) in the loss and metrics, respectively, compared to the single-setting models.}
\label{tab:TV}
{%
\begin{tabular}{@{}lllll|llllc@{}}\toprule
Model & Values & MSE & ESR & RMSE & MSE & ESR & RMSE \\
S6 & & & & & $4.03\cdot 10^{-5}$ & $3.28\cdot 10^{-2}$ & $2.49\cdot 10^{-3}$\\
 & [$a_f$, $r_f$] & $ 3.31\cdot 10^{-5} $ & $ 3.11 \cdot 10^{-2}$ & $ 2.79 \cdot 10^{-3}$ & $ \downarrow 7.33 \% $ & $ \downarrow 7.33 \% $ & $ \downarrow 11.46 \% $\\
& [$a_f$, $r_s$] &  $ 3.11\cdot 10^{-5} $ & $ 4.69\cdot 10^{-2} $ & $ 2.89 \cdot 10^{-3}$ & $ \uparrow 8.08 \% $ & $ \uparrow 8.08 \% $ & $\downarrow 2.95 \% $\\
& [$a_s$, $r_s$] & $ 5.80\cdot 10^{-5} $ & $ 3.35\cdot 10^{-2} $ & $ 3.08\cdot 10^{-3} $ & $ \downarrow 2.91 \% $ & $\downarrow 2.91 \% $ & $\downarrow 10.89 \% $\\
& [$a_s$, $r_f$] & $ 2.40\cdot 10^{-5} $ & $ 1.15\cdot 10^{-2} $ & $ 1.48 \cdot 10^{-3}$ & $\uparrow 68.44 \% $ & $ \uparrow 68.44 \% $ & $ \uparrow 32.84 \% $\\
\hline
S4D & & & & & $4.59\cdot 10^{-5}$ &$ 3.95\cdot 10^{-2}$ & $2.69\cdot 10^{-3}$\\
 &  [$a_f$, $r_f$] & $ 2.95\cdot 10^{-5} $ & $ 2.77\cdot 10^{-2} $ & $ 2.50\cdot 10^{-3} $ & $ \uparrow 12.06 \% $ & $ \uparrow 12.06 \% $ & $\downarrow 3.49 \% $\\
& [$a_f$, $r_s$] & $ 2.82\cdot 10^{-5} $ & $ 4.25\cdot 10^{-2} $ & $ 2.64\cdot 10^{-3} $ & $ \uparrow 66.66 \% $ & $ \uparrow 66.66 \% $ & $ \uparrow 15.57 \% $\\
& [$a_s$, $r_s$]  &  $ 4.31\cdot 10^{-5} $ & $ 2.40\cdot 10^{-2} $ & $ 2.66\cdot 10^{-3} $ & $ \uparrow 50.24 \% $ & $ \uparrow 50.24 \% $ & $ \uparrow 16.69 \% $\\
& [$a_s$, $r_f$]  &$ 3.58\cdot 10^{-5} $ & $ 1.72\cdot 10^{-2} $ & $ 2.10\cdot 10^{-3} $ & $ \uparrow 8.53 \% $ & $ \uparrow 8.53 \% $ & $ \uparrow 3.47 \% $\\
\hline
ED & & & & & $4.84\cdot 10^{-5}$ & $4.07\cdot 10^{-2}$ & $2.86\cdot 10^{-3}$\\
 &  [$a_f$, $r_f$] &  $ 2.91\cdot 10^{-5} $ & $ 2.73\cdot 10^{-2} $ & $ 2.43\cdot 10^{-3} $ & $ \uparrow 6.85 \% $ & $ \uparrow 6.85 \% $ & $ \uparrow 1.47 \% $\\
& [$a_f$, $r_s$]  & $ 2.89\cdot 10^{-5} $ & $ 4.35\cdot 10^{-2} $ & $ 2.68\cdot 10^{-3} $ & $ \uparrow 67.19 \% $ & $ \uparrow 67.19 \% $ & $ \uparrow 18.26 \% $\\
& [$a_s$, $r_s$]  & $ 5.24\cdot 10^{-5} $ & $ 3.03\cdot 10^{-2} $ & $ 3.09\cdot 10^{-3} $ & $ \uparrow 19.6 \% $ & $ \uparrow 19.6 \% $ & $ \uparrow 3.60 \% $\\
& [$a_s$, $r_f$]  & $ 3.53\cdot 10^{-5} $ & $ 1.69\cdot 10^{-2} $ & $ 2.25\cdot 10^{-3} $ & $ \uparrow 45.37 \% $ & $ \uparrow 45.37 \% $ & $ \uparrow 17.03 \% $\\
\hline
LSTM & & & & & $4.64\cdot 10^{-5}$ & $3.63\cdot 10^{-2} $ & $2.76\cdot 10^{-3}$\\
 &  [$a_f$, $r_f$] &$ 4.66\cdot 10^{-5} $ & $ 4.38\cdot 10^{-2} $ & $ 3.15\cdot 10^{-3} $ & $ \downarrow 27.77 \% $ & $ \downarrow 27.77 \% $ & $ \downarrow 21.84 \% $\\
& [$a_f$, $r_s$]  &$ 3.20\cdot 10^{-5} $ & $ 4.83\cdot 10^{-2} $ & $ 2.68\cdot 10^{-3} $ & $ \uparrow 5.06 \% $ & $ \uparrow 5.06 \% $ & $ \uparrow 5.86 \% $\\
& [$a_s$, $r_s$]  &$ 6.91\cdot 10^{-5} $ & $ 3.99\cdot 10^{-2} $ & $ 2.49\cdot 10^{-3} $ & $ \downarrow 11.85 \% $ & $ \downarrow 11.85 \% $ & $ \uparrow 21.42 \% $\\
& [$a_s$, $r_f$]  & $ 2.98\cdot 10^{-5} $ & $ 1.43\cdot 10^{-2} $ & $ 1.76\cdot 10^{-3} $ & $ \uparrow 92.28 \% $ & $ \uparrow 92.28 \% $ & $ \uparrow 55.07 \% $\\
\hline
ED-b & & & & & $1.12\cdot 10^{-4}$ &$ 8.26\cdot 10^{-2}$ & $6.00\cdot 10^{-3}$\\
 &  [$a_f$, $r_f$] & $ 3.12\cdot 10^{-5} $ & $ 2.93\cdot 10^{-2} $ & $ 2.61\cdot 10^{-3} $ & $ \uparrow 112.22 \% $ & $ \uparrow 112.22 \% $ & $ \uparrow 82.02 \% $\\
& [$a_f$, $r_s$]  & $ 3.29\cdot 10^{-5} $ & $ 4.97\cdot 10^{-2} $ & $ 3.05 \cdot 10^{-3}$ & $ \uparrow 116.56 \% $ & $ \uparrow 116.56 \% $ & $ \uparrow 68.75 \% $\\
& [$a_s$, $r_s$]  &$ 1.94 \cdot 10^{-4}$ & $ 1.12\cdot 10^{-1} $ & $ 7.03\cdot 10^{-3} $ & $ \downarrow 40.10 \% $ & $ \downarrow 40.10 \% $ & $ \downarrow 15.95 \% $\\
& [$a_s$, $r_f$]  &$ 1.74 \cdot 10^{-4} $ & $ 8.39 \cdot 10^{-2}$ & $ 7.61\cdot 10^{-3} $ & $ \uparrow 11.48 \% $ & $ \uparrow 11.48 \% $ & $ \uparrow 7.79 \% $\\
\hline
LSTM-b & & & & & $ 2.32\cdot 10^{-4}$ & $1.53\cdot 10^{-1}$& $9.13\cdot 10^{-3}$\\
 & [$a_f$, $r_f$] & $ 3.01\cdot 10^{-5} $ & $ 2.82 \cdot 10^{-2}$ & $ 3.45\cdot 10^{-3} $ & $ \uparrow 37.83 \% $ & $ \uparrow 37.83 \% $ & $ \uparrow 24.98 \% $\\
& [$a_f$, $r_s$]  &  $ 3.77\cdot 10^{-5} $ & $ 5.68\cdot 10^{-2} $ & $ 3.80\cdot 10^{-3} $ & $ \uparrow 167.2 \% $ & $ \uparrow 167.2 \% $ & $ \uparrow 64.34 \% $\\
& [$a_s$, $r_s$]  &$ 8.07 \cdot 10^{-3} $ & $ 4.66\cdot 10^{-2} $ & $ 5.41 \cdot 10^{-2}$ & $ \downarrow 94.18 \% $ & $ \downarrow 94.18 \% $ & $ \downarrow 73.84 \% $\\
& [$a_s$, $r_f$]  &$ 1.85 \cdot 10^{-3} $ & $ 8.89 \cdot 10^{-1} $ & $ 2.83\cdot 10^{-2} $ & $ \downarrow 82.87 \% $ & $ \downarrow 82.87 \% $ & $ \downarrow 58.32 \% $\\
\hline
TCN-b & & & & & $1.59\cdot 10^{-3}$ & $1.19$& $2.04\cdot 10^{-2}$\\
 &  [$a_f$, $r_f$] & $ 1.59 \cdot 10^{-4} $ & $ 1.63 \cdot 10^{-1}$ & $ 5.93\cdot 10^{-3} $ & $ \uparrow 71.24 \% $ & $ \uparrow 71.24 \% $ & $ \uparrow 68.24 \% $\\
 & [$a_f$, $r_s$] & $ 5.64\cdot 10^{-5} $ & $ 8.98 \cdot 10^{-2}$ & $ 4.12\cdot 10^{-3} $ & $ \uparrow 1203.12 \% $ & $ \uparrow 1203.12 \% $ & $ \uparrow 306.6 \% $\\
& [$a_s$, $r_s$] & $ 6.81 \cdot 10^{-3} $ & $ 4.59 $ & $ 4.79\cdot 10^{-2} $ & $ \downarrow 55.58 \% $ & $\downarrow  55.58 \% $ & $ \downarrow 41.37 \% $\\
& [$a_s$, $r_f$]  &$ 3.54 \cdot 10^{-4} $ & $ 1.92 $ & $ 3.33 \cdot 10^{-2} $ & $\downarrow  33.63 \% $ & $\downarrow  33.63 \% $ & $\downarrow  18.7 \% $\\
\end{tabular}}
\end{table*}

\section{CONCLUSION}

The paper proposes a black-box approach to modeling optical dynamic range compressors using a model based on the Selective State Space (S6). Optical compressors are a challenging class of analog audio effects to model due to their nonlinear and time-dependent behavior. These devices' timing characteristics are determined by the pairing of light-emitting and light-sensing components, whose responses depend on both the current and past levels of the input signal. We designed a lightweight architecture that offers minimal latency and computational complexity. This architecture also models the device's variable parameters that influence the compression effect, using Feature-wise Linear Modulation and Gated Linear Units. The architecture featuring the Selective State Space layers has been compared with similar architectures that employ different types of layers commonly used in prior proposals for modeling analog audio effects, such as Long Short-Term Memory (LSTM), Encoder-Decoder (ED), and Structured State Space (S4D) layers. Additionally, state-of-the-art models, including LSTM, ED, and Temporal Convolutional Network (TCN) models, have been incorporated into the comparison.

Results indicated that the S6 model outperforms all other models used in the comparative experiments. Selective State Space layers were shown to capture the devices' temporal characteristics more accurately. Based on quantitative metrics comparing the signals generated by the models with those of the actual devices, these findings are supported by listening tests.

The modeling experiments focused on two optical dynamic range compressors: the Teletronix LA-2A and the TubeTech CL 1B. The LA-2A features variable parameters that determine at what input level compression begins, whereas the amount of compression is controlled by a switch that offers only two modes. In contrast, the CL 1B features four adjustable parameters that govern the onset level of compression, the degree of compression applied, and the attack and release times. The latter case is significantly more challenging than the former, a fact that was reflected in both the objective metrics and the subjective listening evaluations. Additionally, the TubeTech CL 1B dataset presents sparsely sampled control parameters, which adds to the complexity of the modeling task. Nevertheless, the proposed S6 model successfully learned the sound alteration process associated with the control parameters of both devices.

To further analyze the generalization ability of the proposed model, we evaluated its performance in compressing audio signals using parameter combinations not encountered during training. For these experiments, we utilized datasets with varying sampling densities of the control parameters. Results also suggested that denser parameter sampling does not necessarily produce a better generalization capability; instead, there appears to be an optimal balance between too many and too few control parameter sampling points.

The most challenging scenarios involved variable temporal behavior, where datasets of smaller size provided the best results. For the target devices used in this work, the exact parameter values are often not known; they are neither marked around the knobs nor reported in the manuals. Only the extreme values and some intermediate points are typically indicated. We also lack precise knowledge about whether parameter changes occur on a linear or nonlinear scale. Therefore, we opted to sample the parameters using equally spaced intervals, although a more sophisticated sampling strategy could potentially yield better outcomes. 

When evaluating the models' performance on individual compression settings, the combination of fast attack with slow release proved to be the most challenging to emulate. The complexity arises from allowing a very brief initial segment of the sound to pass unaltered before gain reduction begins, which may lead to abrupt dynamic changes within a short time frame. Conversely, when training the models to emulate various compression timing behaviors, there is an expected overall decrease in accuracy. This general reduction in precision is particularly noticeable for settings with slow attack and release times.

Finally, in the conditioning block, we utilize information from the signal's frequency domain based on Fourier analysis. This approach aids in capturing the global properties of a series of input segments, providing information about the input's energy, which influences the compression process. In this regard, alternative frequency domain representations, such as wavelet transforms, could provide a further advantage. They might provide a more effective representation encompassing the frequency and time domains, enabling efficient access to the signal's localized information and potentially enhancing the model's accuracy. 

Further improvements in the model's accuracy could be achieved by designing a loss function that places greater emphasis on errors during sharp input transients, which are particularly challenging to model accurately, especially in the presence of fast attack times. The current loss function, which averages errors across extended temporal segments of the output signal, may be insufficiently sensitive to such errors. Although reducing the training batch size could help address this issue, it would likely lead to a significant increase in training time.

\bibliographystyle{authordate1}
\bibliography{jaes}  

\end{document}